\documentclass[notitlepage,11pt,floatfix,aps,nofootinbib,superscriptaddress]{revtex4-1}

\usepackage{hyperref,bm}

\usepackage{array}

\usepackage{amssymb,amsfonts,multirow}

\usepackage{xcolor}
\usepackage{colortbl}

\definecolor{Gray}{gray}{0.85}
\definecolor{LightGray}{gray}{0.93}
\definecolor{LightGreen}{rgb}{0.88, 1, 0.88}
\definecolor{LightCyan}{rgb}{0.88,1,1}
\definecolor{LightRed}{rgb}{1, 0.85, 0.85}
\definecolor{LightYellow}{rgb}{1, 1, 0.85}
\definecolor{LightBlue}{rgb}{0.87, 0.94, 1}
\definecolor{white}{gray}{1}

\newcolumntype{G}{>{\columncolor{LightGray}}c}
\newcolumntype{L}{>{\columncolor{LightGray}}l}

\usepackage{graphicx}
\usepackage{amsmath}

\usepackage{epsfig}

\usepackage{psfrag}

\usepackage{subfigure}

\newcommand{\bi}{\begin{itemize}}
\newcommand{\ei}{\end{itemize}}
\long\def\del #1 \enddel { }

\def\beq{\begin{equation}}
\def\eeq{\end{equation}}
\def\bea{\arraycolsep .1em \begin{eqnarray}}
\def\eea{\end{eqnarray}}
\def\Tr{{\rm Tr}}

\def\eq#1{(\ref{#1})}

\def\s0#1#2{\mbox{\small{$ \frac{#1}{#2} $}}}
\def\0#1#2{\frac{#1}{#2}}

\usepackage{array,mathtools,amssymb,booktabs}
\newcolumntype{C}{>{$}c<{$}}
\AtBeginDocument{
\heavyrulewidth=.16em
\lightrulewidth=.1em
\cmidrulewidth=.03em
\belowrulesep=.4ex
\belowbottomsep=0pt
\aboverulesep=.4ex
\abovetopsep=0pt
\cmidrulesep=\doublerulesep
\cmidrulekern=.5em
\defaultaddspace=.5em
}

\makeatletter
    \def\CT@@do@color{%
      \global\let\CT@do@color\relax
            \@tempdima\wd\z@
            \advance\@tempdima\@tempdimb
            \advance\@tempdima\@tempdimc
    \advance\@tempdimb\tabcolsep
    \advance\@tempdimc\tabcolsep
    \advance\@tempdima2\tabcolsep
            \kern-\@tempdimb
            \leaders\vrule
                   \hskip\@tempdima\@plus  1fill
            \kern-\@tempdimc
            \hskip-\wd\z@ \@plus -1fill }

   \makeatother

\begin{document}
\title{Global Wilson-Fisher fixed points} 

\author{Andreas J\"uttner} 
\address{
 \mbox{School of Physics and Astronomy, University of Southampton,
 SO17 1BJ, U.K.} }
\author{Daniel F. Litim} 
\address{
 \mbox{Department of Physics and Astronomy, University of
 Sussex, BN1 9QH, U.K.}}
\author{Edouard Marchais} 
 \address{
 \mbox{Department of Physics and Astronomy, University of
 Sussex, BN1 9QH, U.K.}}

\begin{abstract}
The Wilson-Fisher fixed point with $O(N)$ universality in three dimensions  is studied  using the renormalisation group. It is shown how a combination of analytical and numerical techniques determine global fixed points to leading order in the derivative expansion for real or purely imaginary fields with 
moderate numerical effort. Universal and non-universal quantitites such as scaling exponents  and mass ratios are  computed, for all $N$,  together with local fixed point coordinates, radii of convergence,  and parameters which control the asymptotic behaviour of the effective action. We also explain when and why finite-$N$ results do not    converge pointwise towards the exact infinite-$N$ limit. In the regime of purely imaginary fields, a new link between singularities of  fixed point effective actions and singularities of their counterparts by Polchinski are established. Implications for other theories are indicated.
\end{abstract}

\maketitle
\tableofcontents

\section{\bf Introduction}

Wilson-Fisher fixed points provide a paradigm for continuous phase transitions and scale invariance in quantum and statistical field theory \cite{ZinnJustin:2002ru}. 
An important family of Wilson-Fisher fixed points are those of  $O(N)$-symmetric scalar field theories in three dimensions, covering the liquid-vapor phase transitions ($N=1$, Ising model), the ${}^4$He superfluid phase transition ($N=2$, XY model), the ferromagnetic phase transition $(N=3$, Heisenberg model), the QCD phase transition with two massless flavours of quarks ($N=4$), or phase transitions of entangled polymers ($N=0$).  In specific limits such as large-$N$, or close to four dimensions \cite{Wilson:1971dc},
Wilson-Fisher fixed points are under exact perturbative control and allow for a complete solution of the theory \cite{Moshe:2003xn}. At finite $N$, however, Wilson-Fisher fixed points are often strongly coupled and exact analytical solutions are not at hand \cite{Pelissetto:2000ek,Litim:2010tt}. 
Instead, one has to resort to  non-perturbative tools including
numerical simulations on the lattice~\cite{Campostrini:2000iw},
methods from conformal field theory \cite{ElShowk:2012ht} 
or, as will be done here, methods from the renormalisation group~\cite{Litim:2005us,Godina:2005hv,Bervillier:2007rc,Bervillier:2007tc}.

Wilson's renormalisation group is based on the successive integrating-out of momentum fluctuations from a  path-integral representation of the theory 
\cite{Polchinski:1983gv,Wetterich:1992yh,Morris:1993qb,Berges:2000ew,Litim:2000ci}. It allows us
to interpolate between a classical microscopic action at short distances and the full quantum effective action at large distances, supported by powerful optimisation techniques 
\cite{Litim:2000ci,Litim:2001up,Litim:2001fd}. A particular feature of Wilson's continuous RG is its flexibility, permitting systematic and controlled approximations even at strong coupling \cite{Golner:1985fg,Litim:1998nf,Tetradis:1995br,Litim:2002qn,Litim:2002ce,Pawlowski:2003hq}
by combining analytical
\cite{Litim:2002cf,Bervillier:2007rc,Bervillier:2007tc,Bervillier:2008an,
Abbasbandy:2011ij,Litim:2016hlb} 
and numerical tools 
\cite{Adams:1995cv,Bervillier:2007rc,Borchardt:2015rxa}.
Recent applications of the methodology  include 
models of particle physics \cite{Eichhorn:2015kea,Borchardt:2016xju}
and quantum gravity \cite{Litim:2003vp,Benedetti:2012dx,Demmel:2012ub,Dietz:2012ic,Falls:2013bv,Falls:2014tra},
 supersymmetric models \cite{Synatschke:2010jn,Litim:2011bf,Heilmann:2012yf}, 
models in fractional or higher dimensions
\cite{Codello:2014yfa,Percacci:2014tfa,Mati:2014xma,Eichhorn:2016hdi,Kamikado:2016dvw}, 
and models with imaginary couplings \cite{An:2016lni,Zambelli:2016cbw}. 

Ideas to understand interacting fixed points analytically with the help of recursive relations have recently been  developed in \cite{Falls:2013bv,Falls:2014tra,Litim:2016hlb}. The virtue of this approach is that all local couplings, polynomial or otherwise, can be uniquely determined in terms of a few free parameters owing to the iterative diagramatic structure of Wilsonian renormalisation group flows \cite{Litim:2001ky,Litim:2002xm}. 
For $O(N)$ symmetric scalar field theories, exact  recursive relations were obtained and exploited to specify  the  complete fixed point action using local expansions about small, large, or purely imaginary fields \cite{Litim:2016hlb}.  At infinite $N$, it has also been shown that local expansions are sufficient to uniquely determine the  Wilson-Fisher fixed point globally, leading to closed expressions  for the fixed point action. In this limit, free parameters become uniquely specified because the radii of convergence of small and large field expansions overlap. 
In this paper, we are interested in the global Wilson-Fisher fixed point  at finite $N$. The main new ingredient
are the fluctuations of the radial mode which modify the set-up:
local recursive relations are now characterised by a larger number of free parameters, and the singularity structure in the complex field plane is altered. As a result, we observe that
 the 
 radii of convergence for small and large field expansions no longer overlap
 and complete analytical solutions remain out of reach. Still, global fixed points can  be determined accurately, and with moderate  effort, by combining analytical insights with numerical integration \cite{Bervillier:2007rc,Bervillier:2007tc}. Our method identifies the  few free parameters numerically, and offers largely analytical results for the fixed point action for all fields within the respective radii of convergence, together with numerical solutions in those regions of field space which cannot be reached via small or large field expansions. 
 
 Our strategy is put to work for $O(N)$ symmetric scalar field theories 
 covering the universality classes $N=-2,-1,0,1,2,3,\dots,10,20,30,\dots,10^2,10^3$ and $10^4$, and  
to leading order in the derivative expansion. We  compute global fixed point potentials, leading and subleading scaling exponents, mass ratios, local fixed point coordinates, radii of convergence,  as well as the parameters which control local expansions of the effective action. In the regime of purely imaginary fields, a novel relation between singularities of  the Wilson-Polchinski RG  \cite{Polchinski:1983gv} and singularities of their Legendre-transformed counterparts 
\cite{Wetterich:1992yh} is established. Special emphasis is put on the accuracy of numerical procedures and the validity of local expansions, both of which are monitored 
carefully in all steps involved.

The outline of the paper is as follows. Our setup 
and computational strategy is introduced in  Sec.~\ref{sec:RG}, results are detailed in Sec.~\ref{sec:Results}, and
Sec.~\ref{sec:Conclusion} concludes with a discussion and an outlook.

\section{\bf Renormalisation group}\label{sec:RG}

In this section, we shall recall the RG equations, introduce our approximations, 
and explain the computational strategy.

\subsection{Wilsonian renormalisation group}
This work employs the method of functional renormalisation based on a Wilsonian version of the path integral where parts of the fluctuations have been integrated out,   \cite{Wilson:1973jj,Polchinski:1983gv,Wetterich:1992yh,Ellwanger:1993mw,Morris:1993qb, Litim:2000ci}. We consider Euclidean scalar field theories with partition function
\begin{equation}\label{Zk}
Z_k[J]=\int D\varphi\exp(-S[\varphi]-\Delta S_k[\varphi]-\varphi\cdot J)\,,
\end{equation}
where $S$ denotes the classical action and $J$ an external current. The Wilsonian cutoff term  
is given by 
\beq
\Delta S_k[\varphi]=\frac12\int\,\frac{d^dq}{(2\pi)^d}\,\varphi(-q)\, R_k(q^2)\,\varphi(q)
\eeq
with $R_k(q^2\to 0)>0$ for $q^2/k^2\to 0$ and $R_k(q^2)\to 0$ for $k^2/q^2\to 0$ to ensure that $R_k$ acts as an IR momentum cutoff \cite{Litim:2001up,Litim:2000ci,Litim:2001fd}. As such, the partition function \eq{Zk} interpolates between a microscopic (classical) theory $(k\to\infty)$ and the full physical theory $(k\to 0)$.
The main object of our investigation will be the``flowing'' effective action $\Gamma_k$. It relates to \eq{Zk} via a Legendre transformation $\Gamma_k[\phi]=\sup_J(-\ln Z_k[J]+\phi \cdot J) +\Delta S_k[\phi]$, where $\phi=\langle\varphi\rangle_J$ denotes the expectation value of the quantum field. The renormalisation group scale-dependence of $\Gamma_k$ is given by an exact functional identity \cite{Wetterich:1992yh} (see also \cite{Ellwanger:1993mw,Morris:1993qb}),
\begin{equation}\label{FRG}
\partial_t\Gamma_k=\frac12\Tr\frac{1}{\Gamma_k^{(2)}+R_k}\partial_t R_k\,.
\end{equation}
It expresses the change of scale for the effective action $\Gamma_k$ with an operator trace over the full propagator multiplied with the scale derivative of the cutoff itself.  At weak coupling, iterative solutions reproduce the perturbative loop expansion \cite{Litim:2001ky,Litim:2002xm}. The exact RG flow \eq{FRG} also relates to the well-known Wilson-Polchinski flow~\cite{Polchinski:1983gv} by means of a Legendre transformation, and reduces to the Callan-Symanzik equation in the limit where $R_k(q^2)$ becomes a momentum-independent mass term \cite{Litim:1998nf}. The right-hand side (RHS) of the flow \eq{FRG} is  local in field and momentum space implying that the change of $\Gamma_k$ at momentum scale $k$ is governed by fluctuations with momenta of the order of $k$. Optimised choices for the regulator term \cite{Litim:2000ci,Litim:2001up,Litim:2010tt} allow for analytic flows and an improved convergence of systematic approximations \cite{Litim:2005us}. Approximations of the exact flow \eq{FRG} are the central input in this study.

\subsection{Local potential approximation}
The local potential approximation (LPA) expresses the $O(N)$-symmetric 
$3d$ effective action for real scalar fields $\phi$ by a standard kinetic term and a general effective potential,
\begin{equation}\label{Gamma}
\Gamma_k=\int d^3x\left(
	\frac 12 \partial_\mu\vec \phi \partial_\mu \vec\phi + V_k(\phi)
		\right)\,,
\end{equation}		
where $\vec \phi$ denotes a vector of $N$ scalar fields. In the infinite $N$ limit, this approximation becomes exact. 
After inserting \eq{Gamma} into \eq{FRG}, the scale dependence of the dimensionless effective potential
$u(\rho)=V_k(\phi)/k^3$
is determined by the partial differential equation \cite{Berges:2000ew},
\begin{equation}\label{eq:flow}
\partial_t u = -3u+\rho u^\prime +
(N-1)I[u']
+I[u'+2\rho u'']\,.
\end{equation}
Here, $t=\ln k$ is the logarithmic scale parameter, and $k$ the Wilsonian RG scale. The dimensionless field variable
$\rho=\frac 12 \phi^2/k$ automatically
takes into account the invariance of the action under $\phi\rightarrow -\phi$.
The four terms on the RHS originate from the canonical scaling of the potential, the field, the fluctuations of the $N-1$ transversal (Goldstone) modes, and the fluctuations from the longitudinal (radial) mode, respectively. The functions $I[x]$ parametrize  threshold effects owing to  the decoupling of heavy modes. In general, they  take values of order unity for small argument, and vanish in the limit where field-dependent masses $\bar m^2$ grow large ($I[x]\to 0$ for $x\to\infty$, where $x=\bar m^2/k^2$ stands for the square of the field-dependent mass in units of the RG scale).  The functions $I[x]$  also depend  on the choice for the Wilsonian momentum cutoff \cite{Litim:2002cf}. It takes the simple  analytical form
\begin{equation}\label{I}
I[x]=\frac{A}{1+x}
\end{equation}
for an optimised choice of the regulator, following \cite{Litim:2002cf,Litim:2001up,Litim:2000ci}; see \cite{Litim:1995ex,Litim:2016hlb} for exact large-$N$ solutions. The numerical factor $A=2/(d\,L_d)$ arises from the angular integration over loop momenta, and $L_d=(4\pi)^{d/2}\Gamma(d/2)$ denotes the $d$-dimensional loop factor \cite{Litim:2016hlb}. At a fixed point solution, universal scaling exponents are independent of the numerical constant $A$.  This constant can be absorbed into the potential and the fields by the rescaling 
\begin{equation}\label{A}
u\to u/A\,,\quad\rho\to\rho/A\,.
\end{equation}
which is equivalent to  setting $A=1$ in \eq{I}. The benefit of this normalisation is that couplings are now measured in units of the appropriate loop factors 
suggested by naive dimensional analysis~\cite{Giudice:2003tu}. 
All our numerical data below is obtained for this choice. 
Since the zero point energy is not determined by the RG flow \eq{eq:flow}, it is sufficient to investigate the flow of its  first derivative, which is given by
\begin{eqnarray}\label{eq:flowprime}
\partial_t u'  +2u'-\rho u''
&=& -(N-1)\frac{u''}{(1+u^\prime)^2}
-\frac{3u''+2\rho\,u'''}{(1+u^\prime+2\rho u^{\prime\prime})^2}\,.
\end{eqnarray}
We are interested in the Wilson-Fisher fixed point solutions $\partial_t u'=0$ of \eq{eq:flowprime} for all $N$ and all fields.
The fixed point potential obeys an ordinary second order non-linear differential equation
\begin{eqnarray}
2\rho \frac{du''}{d\rho}&=&-\left[3-(N-1)\frac{(1+u'+2\rho u'')^2}{(1+u')^2}\right]\,u''
\label{dglFP}
+(2u'-\rho\,u'')(1+u'+2\rho\,u'')^2\,
\end{eqnarray}
where we recall to have set $A=1$. In principle, \eq{dglFP} uniquely determines the Wilson-Fisher fixed point. In practice, integrating \eq{dglFP} numerically starting from small field values is difficult because the equation is stiff, e.g.~\cite{Morris:1994ki}. Therefore we resort to a combination of analytical and numerical techniques  to identify the unique fixed point solution and its universal properties.

\subsection{Small field expansions}\label{sec:smallfields}
We briefly introduce the polynomial expansions of the scaling
potential around its minimum in a phase with spontaneous symmetry breaking.
It is well-known that scaling exponents can reliably be computed within a small-field expansion \cite{Litim:2002cf}. 
Here we make the polynomial ansatz 
\cite{Margaritis:1987hv,Tetradis:1993ts,Alford:1994fa},
\begin{equation}\label{eqn:pol_approx}
u(\rho) = \sum\limits_{n=2}^{\infty} 
		\frac {\lambda_n}{n!}\left(\rho-\rho_0\right)^n\,,
\end{equation}
which we approximate at a maximum order $n=M$ for the scaling potential in the vicinity
of the potential minimum $\rho_0$ with $u^\prime(\rho=\rho_0)=0$.
We refer to this ansatz as expansion $A$.  This type of ansatz has previously been studied in many works, e.g.~\cite{Litim:2002cf,Aoki:1996fn,Aoki:1998um,Morris:1994ki,Berges:2000ew}.
  After insertion of \eq{eqn:pol_approx} into the flow equation \eq{eq:flow} one obtains a set of coupled ordinary differential equations for the couplings $\lambda_n$ and $\rho_0$. 
 The fixed point conditions become $\partial_t\rho_0=0$ together with
 $\partial_t\lambda_n\equiv\beta_n(\{\lambda_i\})=0$ 
for all $n$. For the expansion $A$, these equations can be solved algebraically,
leading to the exact recursive relation \cite{Litim:2016hlb}
\begin{eqnarray}
\displaystyle
\lambda_{n+2} &=&
\displaystyle
\frac{1}{2\rho_0} 
\bigg( 2(\rho_0\lambda_2 - n) \lambda_{n+1} 
+ (1+2\rho_0\lambda_2) \sum_{k=0}^{n-1} \Omega_{n,k}\sum_{l=0}^{n-k}\Upsilon_{n-k,l}\bigg)
\label{recursiveA}
\end{eqnarray}
where we use the short-hand notation $\Omega_{i,j} =\binom{i}{j}[\delta_j + (2j+1)\lambda_{j+1}+2\rho_0 \lambda_{j+2} ]$ together with  $\delta_j \equiv \delta_{j0}$, and $
\Upsilon_{i,j} = \binom{i}{j} [\delta_{i-j} + \lambda_{i-j+1}]
[\rho_0 \lambda_{j+1} + (j-3)\lambda_{j}]$.
When solved iteratively, the result \eq{recursiveA} provides us with expressions for all couplings $\{\lambda_n,n\ge3\}$ in \eq{eqn:pol_approx} as explicit functions of the vacuum expectation value and the quartic self-coupling,
\begin{equation}\label{para}
\rho_0
\,\ \ {\rm and}\ \ \lambda_2\equiv u''(\rho_0)\,,
\end{equation}
the two couplings which remain undetermined by the recursive solution. 
Results for the   two couplings \eq{para} which remain undetermined by \eq{recursiveA}
 are given below, for all $N$.
The series \eq{eqn:pol_approx} is closely linked to an expansion in the ``symmetric phase''  about vanishing field
 \begin{equation}\label{B}
u(\rho) = m^2\,\rho+\sum\limits_{n=2}^{\infty} 
		\frac {\lambda_{s,n}}{n!}\rho^n\,,
\end{equation}
which we denote as expansion $B$.  Expressions similar to \eq{recursiveA} are found for the fixed point couplings in \eq{B}.
The relevant expressions read 
\cite{Litim:2016hlb}
\bea\label{recursiveN}
\lambda_{s,n+1}
&=& 
\displaystyle
\frac{1}{2n+N} \bigg( (1+m^2) \sum_{l=0}^{n-1} \Lambda_{n,l} 
+ \sum_{k=1}^{n-1} \binom{n}{k} \lambda_{s,n-k+1} \sum_{l=0}^k \Lambda_{k,l} \bigg) \,,
\eea
with $\Lambda_{i,j} = \binom{i}{j} [\delta_{j} + (2j+1)\lambda_{s,j+1}] 
(i-j-3) \lambda_{s,i-j}$. When solved iteratively, the expressions \eq{recursiveN}  provide us with explicit results for all couplings $\{\lambda_{s,n},n\ge2\}$ as functions of the mass term at vanishing field, 
\beq\label{paraB}
m^2\equiv u'(0)\,.
\eeq 
Results for the critical mass parameter \eq{paraB} for all $N$ will be reported below. Notice that \eq{recursiveN} only depends on one unknown parameter \eq{paraB}, whereas \eq{recursiveA} depends on two unknowns \eq{para}. In either case, the remaining parameters must be determined by some other means. However, it has been observed previously that the expansion \eq{eqn:pol_approx} converges more rapidly than expansion \eq{B}, see \cite{Morris:1994ki,Aoki:1996fn,Aoki:1998um,Litim:2002cf,Litim:2003kf}.
For this reason, in the remaining part of this paper, we will focus on the expansion \eq{eqn:pol_approx} with \eq{recursiveA}.
Local polynomial expansions such as \eq{eqn:pol_approx} and \eq{B} necessarily have a finite radius of convergence $R$, 
typically of the order of the vacuum expectation value~\cite{Litim:2002cf,Litim:2016hlb}
\begin{equation}\label{eq:R}
\frac{R}{\rho_0}\approx{\cal  O}(1)\,.
\end{equation} 
In order to extend the fixed point solution for all fields, additional expansions are required.

\begin{table*}
\begin{center}
\begin{tabular}
{cGcGcc}
\toprule
\rowcolor{LightBlue}
&\multicolumn{4}{c}{\bf expansion point in field space}
& \\
\rowcolor{LightBlue}
\multirow{-2}{*}{
${}\ $\bf parameter${}\ $}  
&${}\quad
$minimum$
\quad$
&
${}\
$vanishing field$
\ $&
${}\quad
$large real$
\quad $&${}
$large imaginary$
{}$
&
\multirow{-2}{*}{
${}\quad $\bf info${}\quad$} 
\\
 \midrule
\cellcolor{LightGreen}
&&&&&
\cellcolor{LightGreen}
\\[-3mm]
\cellcolor{LightGreen}
infinite $N$:
&none
&$m^2$
&$\gamma$&$\zeta$
& \cellcolor{LightGreen}
Ref.~\cite{Litim:2016hlb}\\
\cellcolor{LightGreen}
finite $N$:
&$\rho_0$ and $\lambda_2$
&$m^2$
&$\gamma$
&$\zeta$ and $\bar\zeta$
&\cellcolor{LightGreen}
\ \ this work\ \ \\
\bottomrule
\end{tabular}
\caption{The sets of critical parameter (per universality class) which need to be determined to  characterise the Wilson-Fisher fixed point within local expansions
and to leading order in the derivative expansion. Numerical results are summarised in Tabs.~\ref{tab:results_couplings} and~\ref{tab:asymptcoeffs}.}
 \label{tParameters}
\end{center}
\end{table*}

\subsection{Large field expansions}
 To obtain the fixed point solution for all fields, we additionally consider expansions for asymptotically large
real  ($\rho\to\infty$) and imaginary ($\rho\to-\infty$) field amplitudes. In the limit
$\rho\to\infty$, the flow \eq{eq:flowprime}  admits an expansion of the form \cite{Litim:2016hlb}
\begin{equation}\label{eqn:largerho}
	u'(\rho) = \gamma\rho^2\left[1+\sum\limits_{n=1}^\infty\gamma_n\rho^{-n}
		\right]\,,
\end{equation}
which we approximate at some finite maximum order $n=M$.  We refer to this ansatz as expansion $C$. Interestingly,
for all $N$ and using the beta functions for the 
couplings $\gamma_n$, we can solve the fixed point condition recursively to provide us with unique functions for the couplings  \cite{Litim:2016hlb}
\begin{equation}\label{gamma}
\gamma_n=\gamma_n(\gamma)
\end{equation} 
in terms of the leading 
order coefficient $\gamma$,
\beq\label{gamma}
\gamma=\lim_{\rho\to\infty} \s012  u'''(\rho)\,.
\eeq 
Of this one-parameter family of 
solutions, only a single one is related to the Wilson-Fisher fixed point. 
Below, the values \eq{gamma} are determined numerically for each universality class.

We are also interested in the behaviour of the Wilson-Fisher fixed point solution for large negative field squared $\rho\to-\infty$, corresponding to purely imaginary fields. In this limit, the fixed point potential has an expansion of the form \cite{Litim:2016hlb}
\begin{equation}\label{eq:negasympt}
 u'(\rho)=  -1 + \sum\limits_{m=1}^{\infty}\sum\limits_{n=0}^{m-1}
	 \zeta_{m,n}\cdot\left(-\rho\right)^{-m/2}\cdot
	 \left(\ln{\sqrt{-\rho}}\right)^n\,,
\end{equation}
which we approximate at a finite maximal order $m=M$.  We refer to this ansatz as expansion $D$.
In \cite{Litim:2016hlb}, it has been established by exploiting the beta-functions for all couplings $\zeta_{n,m}$ recursively that the putative fixed point couplings are expressed as functions of two independent couplings 
\begin{equation}\label{zetas}
\zeta_{m,n}=\zeta_{m,n}(\zeta,\bar\zeta)\,.
\end{equation}
Of  this two-parameter family of fixed point candidates, only a few, including the Gaussian and the Wilson-Fisher fixed point, extend over all field space. The independent parameters in \eq{zetas} are given by the coefficients
\begin{equation}\label{zetasDef}
\begin{array}{rl}
\zeta&=\zeta_{1,0}\\
\bar\zeta&=\zeta_{4,0}\,,
\end{array}
\end{equation}
see \eq{eq:negasympt}. Values for these will be determined numerically for all $N$.

\subsection{Computational strategy}\label{CS}
A full, global fixed point solution for all values of the field amplitude
$\rho$ and predictions for the
 critical exponents 
can be obtained with
the following two-step strategy. In Step 1, we determine all fixed point couplings in a local approximation, using expansion $A$ owing to its fast convergence. For a given order in the polynomial approximation $M$  
 around the minimum \eq{eqn:pol_approx}, we obtain the set of couplings \eq{recursiveA}.
 Next, we must determine the remaining free parameters \eq{para} numerically. To order $M$ in the polynomial expansion, we impose additional boundary conditions on some of the higher order couplings which are not contained in the approximation up to order $M$. Specifically, we use
\begin{equation}\label{boundary}
\begin{array}{rl}
 \lambda_{M+1}&=0\\[.3mm]
 \lambda_{M+2}&=0
 \end{array}
 \end{equation}
 see \cite{Litim:2002cf,Bervillier:2007rc}, which provides us with two additional equations for the two unknowns \eq{para}. The constraint \eq{boundary} is equivalent to omitting all couplings $\lambda_n$ with $n\ge M+1$ in the effective action from the outset. This strategy has the form of a bootstrap, where \eq{boundary} serves as additional input for closure. Note also that the full non-perturbative fixed point will not obey \eq{boundary} exactly. Hence, it is important to establish stability of our fixed point search under variations in the boundary condition and extensions of the approximation order $M\to M+1$. 
 
The numerical accuracy of the fixed point solution, for all fields,  is monitored by the 
quantity
\begin{equation}\label{Nacc}
 N_{\rm acc}(\rho)=- {\rm log}_{10}\left|\frac{\partial_t u^\prime(\rho)}{u^\prime(\rho)}\right|\,,
\end{equation}
which provides a measure for the deviation from scaling. 
We have $N_{\rm acc}\to\infty$ for the exact solution, for all fields.         
Finally, the universal critical exponent $\nu$ and the sub-leading scaling exponents $\omega_n$
are extracted as the eigenvalues of the stability matrix 
at criticality, 
\begin{equation}
M_{ij}=\left.\frac{\partial\beta_j}{\partial \lambda_i}\right|_{\partial_t u^\prime=0}\,.
\end{equation}
This technique has been successfully applied previously \cite{Litim:2002cf} including for anti-symmetric corrections to scaling \cite{,Litim:2003kf}, and extended for a high-accuracy determination of Ising exponents in
\cite{Bervillier:2007rc}.
Here we adopt it to compute the scaling exponents for all  $N$ ranging between $N=-2$ up to  $N=10^4$.
The accuracy in a scaling exponent $X$ (or a critical coupling) with increasing order in the approximation is 
monitored via
        \begin{equation}\label{NX}
        {N_{X}}=-{\rm log}_{10}\left| 1- \frac{X_n}{X_M}\right|\,.
        \end{equation} 
On the right-hand side, $X_n$ denotes the $n$th order approximation for the 
quantity $X$, and $X_M$ denotes the highest order value thereof. 
Hence, \eq{NX} counts the number of decimal places which agree between the 
approximated and the highest order result. By construction we expect that $X_M$  
approaches the exact solution for $M\to\infty$ with  $N_X\to\infty$.

In order to connect the local fixed point  at small fields with the asymptotic expansions at large fields, we have to determine the values for those parameters which have remained undetermined within the large real \eq{gamma} or large imaginary field solution \eq{zetas},
\begin{equation}\label{unknowns}
\gamma\,, \ \zeta
\,, {\rm and}\ \bar\zeta
\,,
\end{equation}
see Tab.~\ref{tParameters}. Provided the radii of convergence of the small field polynomial expansion $u'_{\rm small}$ and the large field expansion $u'_{\rm large}$ overlap, it is possible to fix the coefficients \eq{unknowns}  in the respective overlapping regimes by requiring $u'_{\rm small}-u'_{\rm large}=0$ to within the desired accuracy. 
In general, we find that the asymptotic expansions for large fields converge, albeit  slowly. While the radii of convergence do overlap in the limit of infinite $N$ \cite{Litim:2016hlb}, this cannot be guaranteed at finite $N$. Furthermore, with decreasing $N$, a very high order in the expansion would be required to potentially achieve 
overlapping radii of convergence with the small field regime and good estimates for the unknown parameters \eq{unknowns}. 
Therefore, as Step 2, we integrate the differential equation \eq{dglFP} at the fixed point numerically to bridge the gap between the small and the large field expansions. This will also provide numerical values for the parameters \eq{unknowns}.

Integrating the stiff differential equation \eq{dglFP} from the vacuum expectation value towards large fields is numerically unstable and can terminate e.g.~in a moving singularity. The instability can be dealt with by integrating into the opposite direction (from large to small fields), which requires a ``shoot and relax"-type strategy. To that end,
we choose field values $\rho_{p1}$ and $\rho_{p2}$ with $0<\rho_{p1}<\rho_{p2}$ and well within the range of 
validity of the small $(\rho_{p1})$ and large field expansion $(\rho_{p2})$, respectively. 
Using initially a trial value for the parameter $\gamma$, we choose the starting point $(u'_{\rm large},u''_{\rm large})$ at $\rho=\rho_{p2}$ to integrate the differential equation  \eq{dglFP} from $\rho_{p2}$ down to $\rho_{p1}$, thereby bridging the two asymptotic expansions. In this direction the integration is stable. The boundary condition for the parameter $\gamma$ at $\rho=\rho_{p2}$ is then adapted iteratively until the result from the numerical integration $u'_{\rm num}$ at $\rho_{p1}$ agrees with the result $u'_{\rm small}$ from the polynomial small field  approximation to within the desired accuracy, say
\beq\label{acc}
|u'_{\rm num}-u'_{\rm small}|<10^{-8}
\eeq 
at the matching point. To ensure stability in the result, we 
varied all matching points over a wide range
and confirmed that our final values for $\gamma$ are independent of the procedure.

In the opposite regime, for large negative $\rho$, we determine
the coefficients of the asymptotic approximation $\zeta_{1,0}$ and
$\zeta_{4,0}$ in a two-step procedure.
We first chose a point $\rho_{m}<0$
for which  the polynomially approximated potential $u'_{\rm small}$ is reliable
with, typically, $N_{\rm acc}(\rho_{m})>10$. 
The starting values $(u'_{\rm small},u''_{\rm small})$ at $\rho_{m}$ 
are used for the numerical integration of \eq{dglFP} towards large negative values $\rho<\rho_{m}$. 
In this direction, the numerical integration is stable.
At sufficiently large negative field values, the leading order behaviour of the solution is solely governed by the coefficient $\zeta_{1,0}$. The sensitivity to the coefficient $\zeta_{4,0}$ is suppressed as $(-\rho)^{-3/2}$. Hence the numerical solution is used to first determine $\zeta_{1,0}$  from
a fit to the asymptotic  expansion at $-\rho\approx 10^{10}$.
We have confirmed that the result is not sensitive against variations of the matching point.

Finally, the value of the subleading coefficient $\zeta_{4,0}$ is  determined by choosing a second matching point
at a significantly smaller purely imaginary field value, together with the numerical result for $u'(\rho)$ and the prior determination of $\zeta_{1,0}$. Again,
we confirm that the result is independent of the details of the procedure, as long as the second matching point is still located within the range of validity
of the asymptotic expansion.  Alternatively, we note that $\zeta_{4,0}$ can also be extracted from the asymptotic expansion of the function $u'+2\rho\,u''$. Here, the dependence on the coefficient $\zeta_{1,0}$ drops out allowing for a  direct access of  $\zeta_{4,0}$ at large negative $\rho$. We have checked that both procedures give the same result.

\section{\bf Results}\label{sec:Results}
This section summarises our results for the scaling solutions, 
the scaling exponents and various other universal and non-universal quantities and their accuracy, the large-$N$ scaling of the theory's parameters, and links with the Wilson-Polchinski RG. 

\subsection{Potential}

At a fixed point of the renormalisation group, quantities are said to be {\it non-universal} provided that their values depend on technical parameters such as the RG scheme. Non-universal quantities cannot be measured in any experiment. This applies to the global shape of scaling potentials and the values of couplings at fixed points, which may vary depending on the choices for the Wilsonian cutoff and the RG scale. Quantities are said to be {\it universal} provided that their values solely depend on the characteristics of the universality class such as the dimensionality of space-time, the short- or long-range nature of interactions, and the number and type of degrees of freedom.  Universal quantities  such as scaling exponents can be measured in experiments.  For the $3d$ $O(N)$ symmetric scalar theories considered here, the different universality classes are then specified by $N$. Finally, quantities are said to be {\it superuniversal} provided they take identical values irrespective of the universality class. 

\begin{figure*}
\includegraphics[width=\hsize]{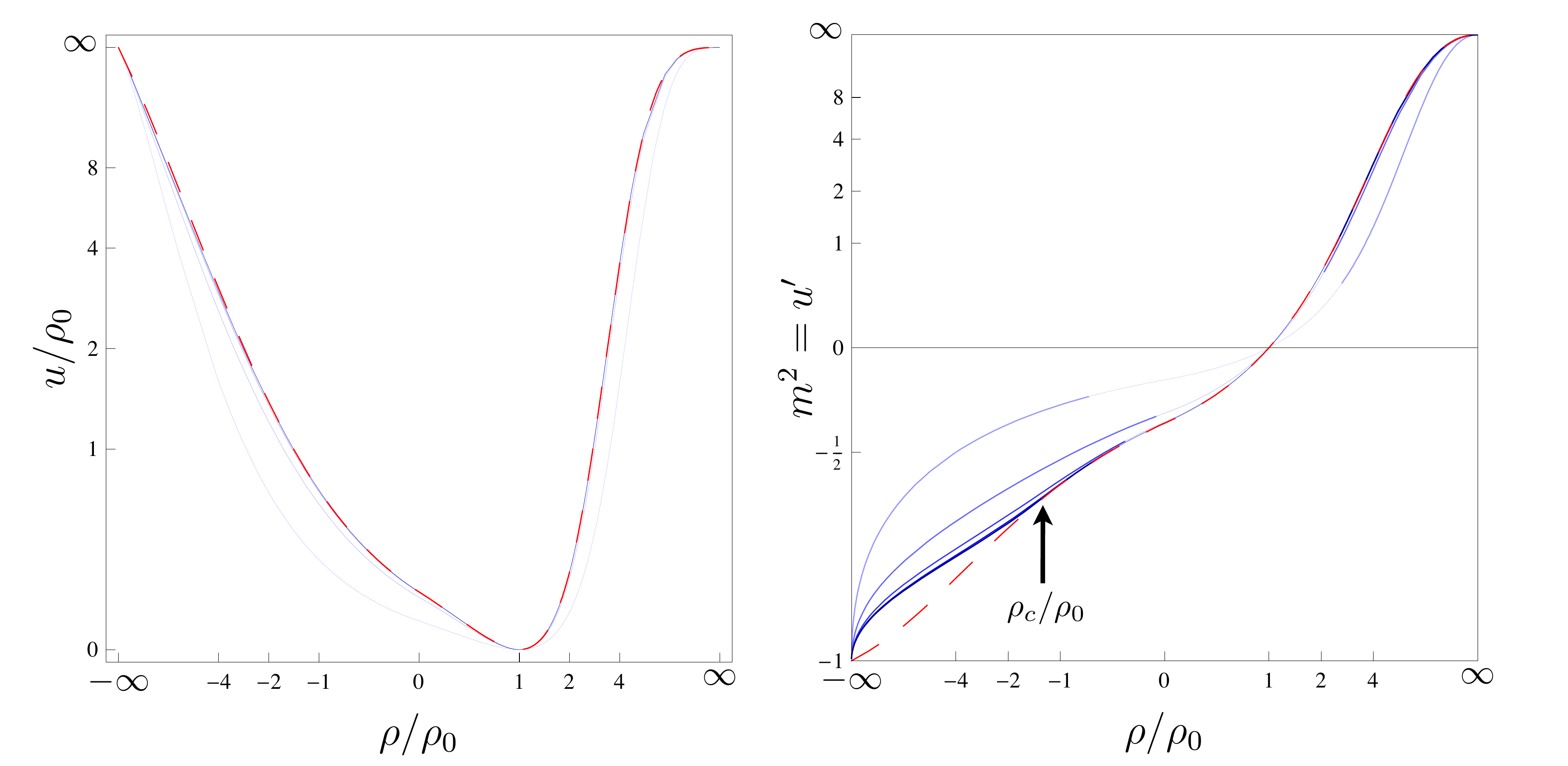}
  \caption{Shown is the potential at the global Wilson-Fisher fixed point solution in units of the vacuum expectation value $u/\rho_0$ (left panel) and its
  first derivative $m^2=u'$ (right panel)  for the universality classes 
	$N=1,10,10^2,10^3$ and $10^4$ (full lines, from light to dark blue). The dashed red curve indicates the infinite-$N$ limit.  The arrow (right panel) indicates the point $\rho_c$ below which finite-$N$ results no longer converge pointwise towards the exact infinite-$N$ result, see \eq{rhoc}. For display purposes, the $x$- and $y$-axes are stretched as $x\to x/(2+|x|)$ and $y\to y/(2+y)$.}
  \label{fig:potential1}
\end{figure*}

\begin{figure*}
\includegraphics[width=\hsize]{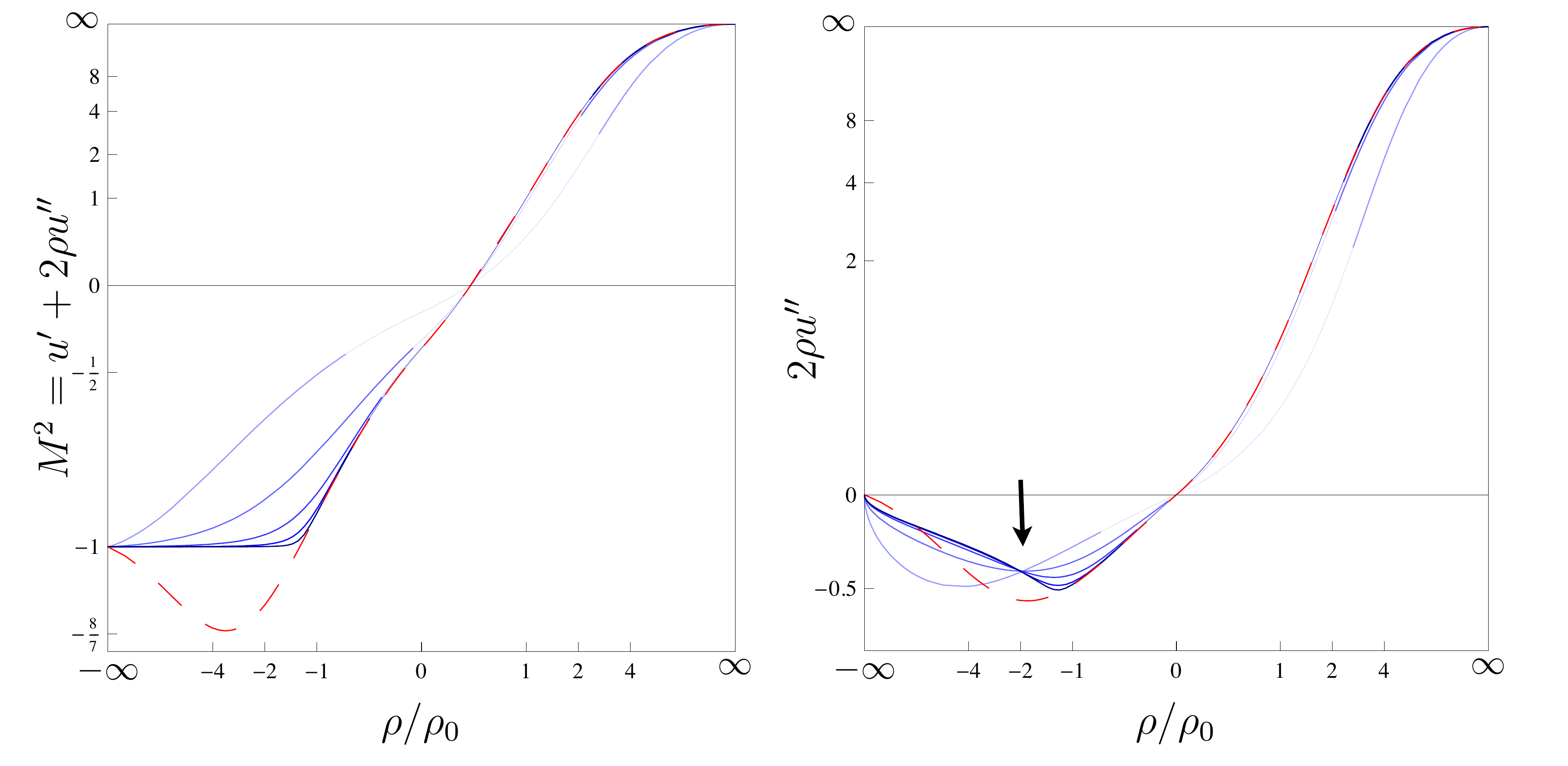}
  \caption{Shown are the field-dependent functions $M^2=u' + 2 \rho u''$  (left panel) and the mass difference $M^2-m^2=2\rho u''$(right panel) at the global Wilson-Fisher fixed point solution for the universality classes 
	$N=1,10,10^2,10^3$ and $10^4$ (full lines, from light to dark blue). The dashed red curve indicates the exact infinite-$N$ limit. The axes are rescaled as in Fig.~\ref{fig:potential1}. The region indicated by the arrow in the right panel 
is studied in more detail in Sec.~\ref{super}.}
  \label{fig:potential12}
 \end{figure*}

Plots of the non-universal global potential and its derivatives are given in Figs.~\ref{fig:potential1},\ref{fig:potential12}
for a representative selection of values for $N$. 
The first plot in Fig.~\ref{fig:potential1} (left panel) shows the results for 
the complete scaling potential $u$ rescaled as in \eq{A} with $A=\rho_0(N)$,
for $N=10^n, n=0,1,2,3$ and $4$ 
(the intensity of the line colour increases in this order). 
Notice that the $x$- and $y$-axes are rescaled as $x\to x/(2+|x|)$ and $y\to y/(2+y)$ to allow a display of the global fixed point potential for all real and imaginary fields. 
The potentials show a single global minimum at $\rho_0$ for all $N$ and 
all fields. 
The plots at finite $N$ are complemented by the  solution in the
 infinite-$N$ limit (red dashed line, cf.~\cite{Litim:2016hlb}). With increasing but finite $N$, the solutions seem to approach the infinite-$N$ result smoothly (note that the red dashed
line is nearly on top of the large-$N$ solution for $u/\rho_0$ for 
all $\rho$ and is therefore hardly visible) -- see Sec.~\ref{vs} for a detailed discussion of the $N$-dependence. 
Our results can further be used to derive the global equation of state and universal amplitude ratios for all $N$  (for an overview, see \cite{Pelissetto:2000ek}),  following the lines of \cite{Seide:1998ir} for the Ising universality class at LPA and beyond (e.g.~including  corrections through anomalous dimensions).


\subsection{Mass scales}\label{mass}
Next we discuss derivatives of the scaling potential. The RG flow is driven by the two relevant mass scales of the model, the field-dependent Goldstone mass $\bar m^2\equiv m^2\,k^2$ and the field-dependent mass of the radial mode $\bar M^2\equiv M^2\,k^2$ which, in units of the RG scale, are related to the potential as
\beq\label{mM}
\begin{array}{rcl}
m^2&=&u'(\rho)
\\
M^2&=&u'(\rho)+2\rho\,u''(\rho)\,.
\end{array}
\eeq
Either mass term contributes to the flow except at infinite $N$ $(N=1)$ where the radial (Goldstone) contribution is absent. As can be seen from the flow equation \eq{eq:flowprime} together with \eq{mM},
the limits 
\begin{equation}\label{convexity}
m^2,M^2\to -1
\end{equation}
signal a putative singularity \cite{Tetradis:1992qt,Litim:2006nn}. Dynamically, however, the limit \eq{convexity} is achieved smoothly without encountering any singularity, as follows from  Figs.~\ref{fig:potential1},\ref{fig:potential12} and from the analytical result \eq{eq:negasympt}, see \cite{Litim:2016hlb}. 
Also, the mass scale ordering $m^2<M^2$ for large positive $\rho>0$ becomes inverted  $-1<M^2\le m^2$ once $\rho<0$. With increasing $N$, the radial mass fully dominates the solution for imaginary fields.

\begin{figure}[t]
   \includegraphics[width=.8\hsize]{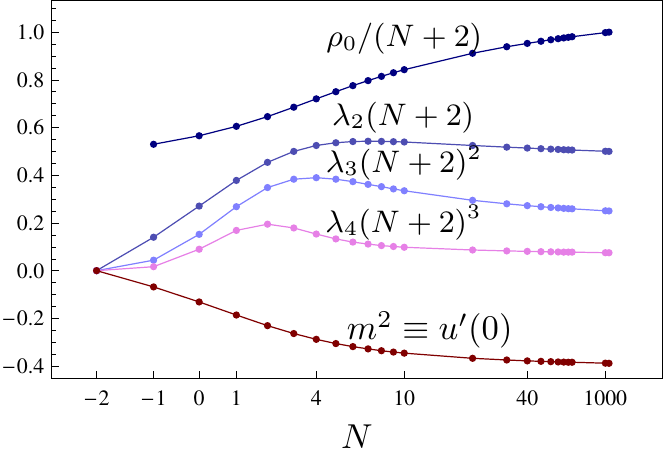}
 \caption{Shown is the $N$-dependence of couplings characterising the Wilson-Fisher fixed point. These include
the dimensionless vacuum expectation value $\rho_0(N)$, the mass term squared at vanishing field $m^2=u'(0)$,  and the first few polynomial couplings $\lambda_{2-4}$ at the potential minimum. 
\label{fig:scaling_coeffs}}
\end{figure}

\begin{figure}[t]
   \includegraphics[width=.8\hsize]{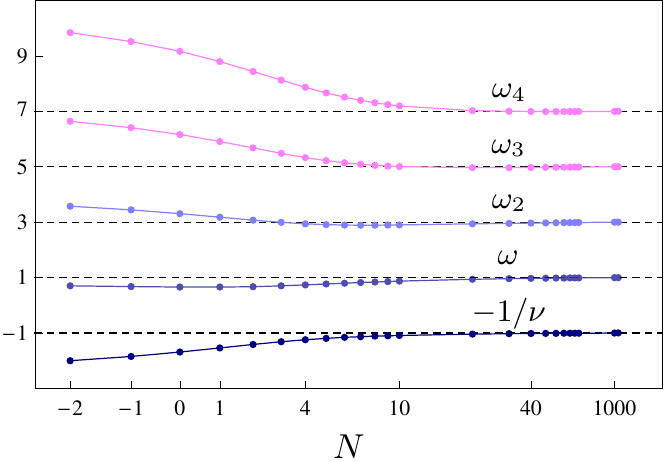}
 \caption{Shown are the first few universal  
	eigenvalues $-1/\nu$, $\omega$ and $\omega_{(2,3,4)}$ for all $N$, in comparison with the large-$N$ result (dashed horizontal lines).\label{fig:scaling_coeffs2}}
\end{figure}

\subsection{Finite vs infinite $N$}\label{vs}
In  Figs.~\ref{fig:potential1}, \ref{fig:potential12}, results at finite $N$ are compared with exact solutions at infinite $N$. With increasing $N$, the global effective potential appears to approach the infinite-$N$ result smoothly. On a finer level, however, we observe that the masses \eq{mM} do not converge pointwise to the infinite-$N$ result.  This is evidenced by the data for the field-dependent masses  $m^2$, $M^2$, and their difference  $M^2-m^2$ as shown in Figs.~\ref{fig:potential1}, \ref{fig:potential12}. 
The dashed red line representing the solution in the infinite $N$ limit differs notably from the solutions at any finite $N$ for $\rho<\rho_c$.
The strict convexity bound \eq{convexity} is relaxed for the radial modes at infinite $N$, admitting $M^2<-1$ without causing a singularity along the flow. At finite $N$, the radial mode fluctuations contribute to the flow and the convexity bound $M^2\ge -1$  must be  adhered to.
Quantitatively, an estimate for $\rho_c$ is found from resolving $M^2=-1$ for $\rho$ at infinite $N$, leading to
\beq\label{rhoc}
\begin{array}{rcl}
\rho_c/\rho_0&=&-1.2687\cdots
\\[.5ex]
u'(\rho_c)&=&-0.64867\cdots\,.
\end{array}
\eeq
Notice that  \eq{rhoc} is  consistent with the radius of convergence $R_D\approx -1.5$  determined from the expansion \eq{eq:negasympt} at infinite $N$ \cite{Litim:2016hlb}.
For fields below \eq{rhoc}, $\rho<\rho_c$, the infinite-$N$ solution no longer fullfills $M^2\ge -1$. Hence, \eq{rhoc} denotes the smallest possible value for $\rho$ down to which finite-$N$ results may converge pointwise  towards the infinite~$N$ expressions. 
Furthermore, from the data, we also observe that the scaling solutions do approach the infinite $N$ result for all fields with $\rho>\rho_c$.  Taking the limit $N\to\infty$ for the solutions of the finite-$N$ RG equations  \eq{eq:flowprime}, we observe that $M^2\to -1$ in the entire regime $\rho<\rho_c$. This is the convex hull of all functions $M^2$ for any finite $N$, defined via a differential equation for $u'$ which is solved by 
\beq\label{convexhull}
u'=-1+\frac{\zeta}{\sqrt{-\rho}}
\eeq
for all $\rho<\rho_c$. The coefficient $\zeta$ is given by $\zeta_{1,0}$ in \eq{eq:negasympt}, in the limit $N\to\infty$. We observe the absence of subleading terms in  \eq{convexhull}, indicating that all subleading coefficients $\zeta_{m,n}$ with $m\ge 2$ in the general result \eq{eq:negasympt} become suppressed parametrically for  sufficiently large $N$. 

We  conclude that the large-$N$ limit of the flow  \eq{eq:flowprime} at finite $N$ does not commute  for all fields with the integration of the flow at infinite $N$. Rather, this limit is 
qualitatively different from the solution of the large-$N$ equations in the regime  for purely complex fields.
The culprit for these differences are the flucutations  of the longitudinal (radial) mode, which at any finite $N$  interfere with those of the transversal (Goldstone) modes and modify  the analytic structure of the fixed point solution.

\begin{figure}[t]
 \includegraphics[width=.8\hsize]{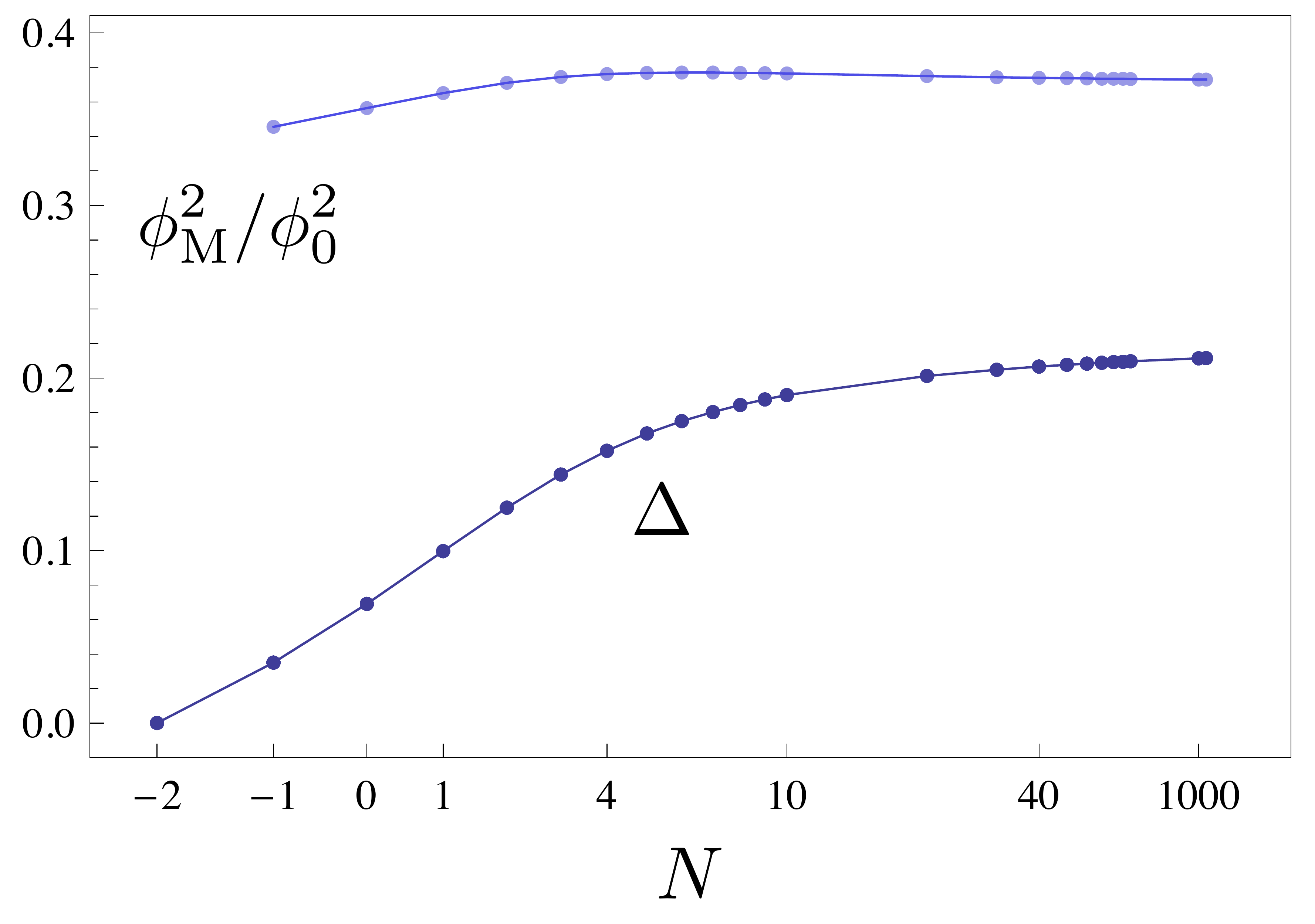}
 \caption{The ratio of the field value at the inflection point 
 	and at the potential minimum $\phi^2_{\rm M}/\phi^2_0$ \eq{ratio}, and the potential barrier $\Delta$ \eq{deltau} as functions of the universality class $N$ (see main text).}\label{barrier}
\end{figure}

\subsection{Mass differences}\label{super}

Next, we investigate in more detail the region indicated by the arrow in Fig.~\ref{fig:potential12} (right panel), where the field-dependent mass difference $
Y\equiv M^2-m^2$ is displayed as a function of $X\equiv \rho/\rho_0$. We use the vacuum expectation value $\rho_0$ to set the ``scale'' in field space, with $X$ measuring the magnitude of fields in units of the vev, per universality class. 
Although the functions $M^2$ and $m^2$ take quite  different shapes for different $N$,
we observe from the data that the functions $Y(X)$ seem to coincide close to the region where $X\approx -2$, Fig.~\ref{fig:potential12}. An exact coincidence for all $N$ would be an indication for superuniversal behaviour. 
However,
a closer inspection shows that this is not quite the case:
for each pair of universality classes $(N_1,N_2)$ with $N_1\neq N_2$ we have computed the field ratio $X$ such that
$Y(N_1)=Y(N_2)$, leading to a collection of coincidence points $(X,Y)$. For large $N_1$ and $N_2$, the data points show a  mild
dependence on $N$ with values in the range $X\approx [-1.7,-2.2]$ and $Y\approx [-0.27,-0.28]$.
We conclude that within the approximations used here
the mass difference does not show superuniversal behaviour other than the trivial one where $(X,Y)=(1,0)$ for any $N$, reflecting  equality of masses at vanishing field.
Note also that we have excluded the exact infinite $N$ result from the analysis, because the pointwise convergence of finite-$N$ results  towards their large-$N$ limit no longer coincide with the exact infinite $N$ result for the range of $X$ values found above ($X<\rho_c/\rho_0$), see Sec.~\ref{vs}. 
It would be interesting to check corrections to this picture beyond the local potential approximation.

\subsection{Field ratios}

At the Wilson-Fisher fixed point, the physical masses \eq{mM} vanish identically in the infra-red limit $k\to 0$. Furthermore, away from the fixed point both masses always coincide for vanishing field, $m^2(\rho=0)=M^2(\rho=0)<0$. At $k\neq 0$, and with increasing field values, both masses increase, with $M^2>m^2$ for all fields $\rho>0$. 
The square of the Goldstone mass changes sign at the potential minimum, $\rho=\rho_0\equiv\frac12\phi^2_{0}/k^2$   
and the square of the radial mass changes sign at the inflection point $\rho=\rho_M\equiv\frac12\phi_M^2/k^2$.
This makes their ratio
\begin{equation}\label{ratio}
\rho_M/\rho_0\equiv {\phi^2_M}/{\phi^2_{0}}
\end{equation}
a potentially universal quantity worth being explored, also in the light of our results in the previous section. The ratio also gives access to global aspects of the fixed point solution. At the Wilson-Fisher fixed point in the infinite $N$ limit, we find the ratio of field values at the inflection point and at the potential minimum analytically. It reads
\begin{equation}
{\phi^2_M}/{\phi^2_{0}}=\frac{1}{5}\frac1{(1+x)^2}\,,
\end{equation}
where $x=-0.26763\cdots$ is determined as the unique solution of the transcendental equation $0=8+25 x +15x^2+15\sqrt{x}(1+x)^2\arctan\sqrt{x}$, leading to
\begin{equation}\label{ratioInf}
{\phi^2_M}/{\phi^2_{0}}=0.37288\cdots\,.
\end{equation}
Our numerical results for finite $N$ are displayed in Fig.~\ref{barrier}. It is noteworthy that the asymptotic limit \eq{ratioInf} is an excellent approximation for \eq{ratio} for most $N$ down to $N\approx 4$, suggesting that \eq{ratio} is largely super-universal, $i.e.$~independent even of the universality class.  Quantitatively, the $N$-dependence of \eq{ratio} is much weaker than the $N$-dependence observed for the universal scaling exponents themselves, and much less so than the fixed point coordinates.

\subsection{Barrier height}
In Fig.~\ref{barrier} we also display the ``barrier height'' --- refering to the difference between the potential at vanishing field and at the global minimum  of the fixed point solution --- as a function of the universality class and in units of the vacuum expectation value
\begin{equation}\label{deltau}
\Delta=\frac{u(0)-u(\rho_0)}{\rho_0}\,.
\end{equation}
The dimensionless parameter \eq{deltau} supplies us with global information of the fixed point potential. It is of order unity for all $N$, and characterises the universality class. Note that the physical barrier height is given by $k^3(u(\rho_0)-u(\rho))$, and vanishes in the infrared limit $k\to 0$ as it must at a second order phase transition.  In the infinite-$N$ limit, we have
\begin{equation}
\Delta=-\frac13\frac{x}{1+x}
\end{equation}
where $x=-0.388347\cdots$ uniquely solves the trans\-cendental equation 
$H(x)=-1$, with
\begin{equation}\label{H}
H(x)=\032\sqrt{x}\arctan\sqrt{x}+\012\0x{1+x}\,.
\end{equation}
Quantitatively, we therefore have
\begin{equation}
\Delta=0.211638\cdots\,.
\end{equation}
As can be seen from Fig.~\ref{barrier}, the $N$-dependence of \eq{deltau} is much more pronounced than the one for the scaling exponents and for the field ratio, in particular for small $N$.

\begin{table*}
\begin{tabular}{lllllll}
\rowcolor{LightGreen}
\toprule
&&&&&&\\[-3mm]
\rowcolor{LightGreen}
$\bm N$&$\bm M$&$\bm{\rho_0}$&$\bm{m^2\equiv u^\prime(0)}$&$\bm{\lambda_2}$&$\bm{\lambda_3}$&$\bm{\lambda_4}$\\
\midrule
&&&&&&\\[-3mm]
$-2$&33&0   &0&0.139993154&0.0475919230&0.0211562710\\
\rowcolor{LightGreen}
\toprule
&&&&&&\\[-3mm]
\rowcolor{LightGreen}
$\bm N$&$\bm M$&$\bm{\rho_0/(N+2)}$&$\bm{m^2\equiv u^\prime(0)}$&$\bm{\lambda_2(N+2)}$&$\bm{\lambda_3(N+2)^2}$&$\bm{\lambda_4(N+2)^3}$\\
\midrule
&&&&\\[-3mm]
$-1$&33&0.53010731451&$-$0.0683026035970&0.1398353693&0.04430604100&0.01677868580 \\
\rowcolor{LightGray}
0&33&0.5654460144&$-$0.131345164893&0.2706917649&0.1520450814&0.0898543170\\
1&33&0.60496613456&$-$0.186064249476&0.3784932636&0.2683346829&0.1690960425\\
\rowcolor{LightGray}
2&33&0.645985012570&$-$0.2301854999&0.45440714932&0.34808205451&0.19512012101\\
3&33&0.6851667859&$-$0.263517269&0.50020343745&0.383421958193&0.178857298500\\
\rowcolor{LightGray}
4&33&0.7201520549&$-$0.28778463&0.52455774613&0.390008902143&0.153420501328\\
5&31&0.7501424439&$-$0.3053458&0.53621216879&0.383637436049&0.133374923843\\
\rowcolor{LightGray}
6&31&0.7753969531&$-$0.3182375&0.54105444077&0.373076375122&0.119991941052\\
7&31&0.7965902702&$-$0.3279242&0.542409476345&0.362059820747&0.111277257677\\
\rowcolor{LightGray}
8&33&0.814447579888&$-$0.3353884&0.542027337724&0.351907477606&0.105437978454\\
9&33&0.829605953766&$-$0.34127835&0.540810681071&0.342956872311&0.101340258191\\
\rowcolor{LightGray}
10&33&0.842584531609&$-$0.3460262&0.539224887998&0.335185982691&0.0983242065590 \\
20&33&0.911388350304&$-$0.36745406&0.525086952003&0.294955212949&0.0866005267390 \\
\rowcolor{LightGray}
30&33&0.938518498813&$-$0.374511202&0.517868217096&0.280245472539& 0.0828790249620\\
40&33&0.952953265258&$-$0.378008299&0.513816036847&0.272755783975& 0.0809771159500\\
\rowcolor{LightGray}
50&33&0.961904553341&$-$0.3800950613&0.511248174538&0.268231429777& 0.0798166118300\\
60&33&0.967996091673&$-$0.3814811685&0.509480655859&0.265205203302& 0.0790338619260\\
\rowcolor{LightGray}
70&33&0.972408915311&$-$0.3824686839&0.508191322583&0.263039546849& 0.0784700610060\\
80&33&0.975752675739&$-$0.38320789159&0.507209817973&0.261413388856& 0.0780445668020\\
\rowcolor{LightGray}
90&33&0.978373784195&$-$0.383781972477&0.506437902150&0.260147596608& 0.0777120320860\\
$10^2$&33&0.980483612096&$-$0.38424069090&0.505815034989&0.259134394528& 0.0774449877720\\
\rowcolor{LightGray}
$10^3$&31&0.998004906308&$-$0.387937969253&0.500598177540&0.250914244992& 0.0752479053650\\
$10^4\ \ $&$31\ \ $&$0.999800049134\quad$&$-$$0.38830586281\quad$&$0.500059981802\quad$&$0.250091428207\quad$& $0.0750248247680$\\
\midrule
\rowcolor{LightYellow}
 &&&&&&\\[-3mm]
 \rowcolor{LightYellow}
$\infty$&$\infty$&\quad\quad\quad1
&$-0.388 346 718 912 \cdots$&\quad\quad\quad$\frac12$&\quad\quad\quad$\frac14$&\quad\quad\quad$\frac3{40}$\\[.5mm]
\bottomrule
\end{tabular}
\caption{Numerical results for the Wilson-Fisher fixed point for all $N$ parametrised in terms of 
the (dimensionless) vacuum expectation value $\rho_0$, the first few polynomial couplings at the potential minimum, and the mass term squared at vanishing field. For $N\neq-2$, couplings have been scaled with appropriate powers of $(N+2)$. The quoted digits are significant
	according to the criterion detailed in Sec.~\ref{sec:exponents}. We have 
	cut the number of quoted digits at a maximum of 12.  
	}\label{tab:results_couplings}
\end{table*}

\subsection{Polynomial couplings} \label{sec:exponents}
The coordinates of the vacuum expectation value $\rho_0$, the dimensionless mass parameter at the potential minimum $m^2=u'(0)$, and the polynomial fixed point couplings $\lambda_i$  are summarised in 
Tab.~\ref{tab:results_couplings} (see also Fig.~\ref{fig:scaling_coeffs}).
We restrict ourselves to presenting only the 
leading and the first few  subleading couplings and  recall that perturbative loop factors have been scaled into the couplings. 
The order $M$ up to which we approximate the expansion around the minimum, is 
also given in the table. The quoted digits are those which did not
change for at least four preceeding orders of the approximation (that is, between  $M-4$ and $M$). 
\begin{table*}
\begin{tabular}{lllllll}
\toprule
\rowcolor{LightGreen}
 &&&&&&\\[-3mm]
\rowcolor{LightGreen}
$\bm N$&$\bm M$&\quad\quad\quad$\bm{\nu}$
&\quad\quad\quad$\bm{\omega}$
&\quad\quad\quad$\bm{\omega_2}$
&\quad\quad\quad$\bm{\omega_3}$\\
\midrule
$-$1&33&0.5415364745&0.6757184&3.4431&6.4099\\
\rowcolor{LightGray}
0&33&0.5920826926&0.65787947&3.3084&6.1631\\
1&33&0.6495617738&0.65574593&3.18001&5.91223\\
\rowcolor{LightGray}
2&33&0.70821090748&0.671221194&3.071402&5.67904\\
3&33&0.7611231371&0.6998373178&2.9914230&5.4826528\\
\rowcolor{LightGray}
4&33&0.804347696&0.733752926&2.9399940&5.330637\\
5&31&0.8377407110&0.76673529&2.9108908&5.219539\\
\rowcolor{LightGray}
6&31&0.8630761595&0.795814494&2.896726&5.140977\\
7&31&0.8823889567&0.820316404620&2.8916166&5.086305\\
\rowcolor{LightGray}
8&33&0.897337664625&0.8406122805&2.89163472&5.04848125\\
9&33&0.909128139450&0.85738397045&2.89438031&5.0223239\\
\rowcolor{LightGray}
10&33&0.918605123154&0.87131097659&2.898458278&5.00419855\\
20&33&0.960678346035&0.936742371978&2.93751369319&4.96566025423\\
\rowcolor{LightGray}
30&33&0.974173017876&0.958441374389&2.95672986621&4.97022378261\\
40&33&0.980781323265&0.969097610936&2.96709100884&4.97540746495\\
\rowcolor{LightGray}
50&33&0.984698956658&0.975413687073&2.97348964704&4.97932916821\\
60&33&0.987290505905&0.979589069170&2.97781818967&4.98225142783\\
\rowcolor{LightGray}
70&33&0.989131594889&0.982553360520&2.98093683915&4.98447973715\\
80&33&0.990506893951&0.984766383156&2.98328905465&4.98622431780\\
\rowcolor{LightGray}
90&33&0.991573286057&0.986481464992&2.98512577783&4.98762303949\\
$10^2$&33&0.992424323549&0.987849597646&2.98659941340&4.98876757415\\
\rowcolor{LightGray}
$10^3$&31&0.999249240822&0.998798467576&2.99865080272&4.99880711432\\
$10^4$\quad\quad&31\quad\quad&0.999924992406\quad\quad &0.999879984650\quad\quad&2.99986500784\quad\quad&4.99988007059\quad\quad\\
\midrule
\rowcolor{LightYellow}
 &&&&&&\\[-3mm]
\rowcolor{LightYellow}
$\infty$&$\infty$&\quad\quad\quad 1&\quad\quad\quad 1&\quad\quad\quad 3&\quad\quad\quad 5\\
\bottomrule
\end{tabular}
\caption{Numerical results for the scaling exponents  for all 
	universality classes considered. All quoted digits are significant
	according to the criterion detailed in Sec.~\ref{sec:exponents}. For some indices the achieved accuracy reaches about 30 digits. For display purposes, we have 
	cut the number of quoted digits at a maximum of 12.
 	}\label{tab:results_scaling}
\end{table*}
In the last row of Tab.~\ref{tab:results_couplings} we also
provide the $N\to\infty$ limits, empirically determined 
for the couplings $\lambda_i$ based on \eq{eqn:largeN} 
in the case of the scaling exponents, and based on~\cite{Litim:2016hlb}
for $m^2$.
For $N=-2$ the potential's minimum resides exactly at the origin, $\rho_0=0$. As $N$ increases
$\rho_0$ moves towards larger values, asymptotically reaching $\rho_0=N$ for $1/N\to 0$. Hence, $\rho_0(N)$ sets a ``scale'' for the overall normalisation of couplings with $N$.
For this reason, it is  natural to use $\rho_0(N)$ as a normalisation factor for the results. For the purpose of Tab.~\ref{tab:results_couplings} we therefore have rescaled all couplings with (appropriate powers of) $N+2$. In this way  all couplings achieve a finite limit even for large $N$.\footnote{Note that this rescaling is equivalent to the rescaling \eq{A} of the flow equation \eq{eq:flowprime} with $A=N+2$.} By definition, the scaled couplings  vanish in the limit $N\to -2$. Quartic couplings are of order unity. By naive dimensional analysis \cite{Giudice:2003tu}, the fixed point is boarderline strongly coupled for most $N$. We also observe that couplings reach maxima for moderate values of $N$, and that they approach their finite large-$N$ limits from above.

\subsection{Exponents}\label{Exponents}
Exact results for the scaling exponents in $3d$ are known in the 
infinite-$N$ limit~\cite{Stanley:1968gx,Litim:2001dt} where
the universal eigenvalues $\theta$ approach those of the spherical model. One finds
\begin{equation}\label{eqn:largeN}
\theta_n=2n-1+O(1/N)\,,
\end{equation}
where $n=0,1,2,\cdots$ takes integer non-negative values. The sole negative eigenmode is related to the exponent $\nu$ by $\nu=-1/\theta_0$, and the sub-leading corrections-to-scaling exponents are $\omega_n=\theta_n$.
In the vicinity of $N=-2$, the exact result reads~\cite{Fisher:1973zzb,Litim:2001dt}
\begin{equation}
\nu=\frac12+O(N+2)\,.
\end{equation}
Numerical results for exponents at finite $N$ have been given in \cite{Litim:2001fd,Litim:2002cf} with an accuracy up to six significant digits to leading order in the derivative expansion.\footnote{See \cite{Litim:2010tt} for an overview of results up to fourth order in the derivative expansion.}  In \cite{Bervillier:2007rc}, the accuracy has been extended up to 14 significant digits for the Ising universality class. The present study provides the exponents for any $N$ within $-2<N<\infty$, and enhances the accuracy up to about 30 digits in the result. Our numerical results all smoothly approach
their known asymptotic value as $N$ is increased. This is illustrated 
in Figs.~\ref{fig:scaling_coeffs}, \ref{fig:scaling_coeffs2}. The data underlying these plots is 
summarised in  Tab.~\ref{tab:results_scaling}.
It is worth pointing out that the Wilsonian regulator function $R_k$ in \eq{FRG} is key  to help enhancing convergence and stability of results \cite{Litim:2002cf,Litim:2001up,Litim:2001dt,Litim:2000ci}.  Moreover, universal  exponents show a characteristic ``cusp-like'' structure as a function of the RG scheme, with best results located exactly at the cusp  \cite{Litim:2007jb}. Incidentally,  cusp-like structures have also been observed within the conformal boostrap technique \cite{ElShowk:2012ht}.
 For our models, the cusp is achieved for the optimised cutoff $R_k\sim (k^2-q^2)\theta(k^2-q^2)$  \cite{Litim:2001up}. Hence, results are ``as good as it gets'' to leading order in a derivative expansion, in exact agreement with results from Polchinski's equation \cite{Polchinski:1983gv,Litim:2005us}.  

 For completeness, we compare our findings to leading order in a derivative expansion with the best estimates to date based on Monte Carlo (MC) simulations and the conformal bootstrap where impressive levels of accuracy have been achieved as of late  \cite{PhysRevE.94.052102,Hasenbusch:2011yya,Kos:2016ysd,Campostrini:2006ms}. Specficially, Ising exponents $\nu=0.629971(4)$ and $\eta= 0.0362978(20)$ have been found recently with the conformal bootstrap technique   \cite{Kos:2016ysd}.   MC simulations have found  $\nu=0.58759700(40)$ for entangled polymers ($N=0$) \cite{PhysRevE.94.052102}; $\nu=0.63002(10)$ and $\eta= 0.03627(10)$ for Ising universality  $(N=1)$   \cite{Hasenbusch:2011yya}, and $\nu=0.6717(1)$ and $\eta = 0.0381(2)$ for the XY model ($N=2$) \cite{Campostrini:2006ms}.
  LPA results for $\nu$ agree with these to within 0.75\% ($N=0)$, $3\%\, (N=1)$, and $5\%\, (N=2)$, respectively. Subleading scaling exponents such as $\omega$ \cite{Litim:2001dt} or antisymmetric corrections to scaling $\omega_5$ \cite{Litim:2003kf} are more sensitive to higher derivative interactions \cite{Litim:2010tt} and deviate more strongly in LPA \cite{Litim:2001dt} (often in the 25\%-40\% range) provided they are correlated with $\nu$ (if not, much smaller deviations are observed \cite{Litim:2001dt}).  
Sub-percent  accuracy  of RG results is achieved through suitably ``optimised'' higher orders in the derivative and/or vertex expansion; see \cite{Canet:2002gs,Canet:2003qd,Canet:2004xe,Litim:2010tt} for results up to fourth order in the derivative expansion.

\subsection{Large-field asymptotics}

\begin{figure}
\begin{center}
\includegraphics[width=.8\hsize]{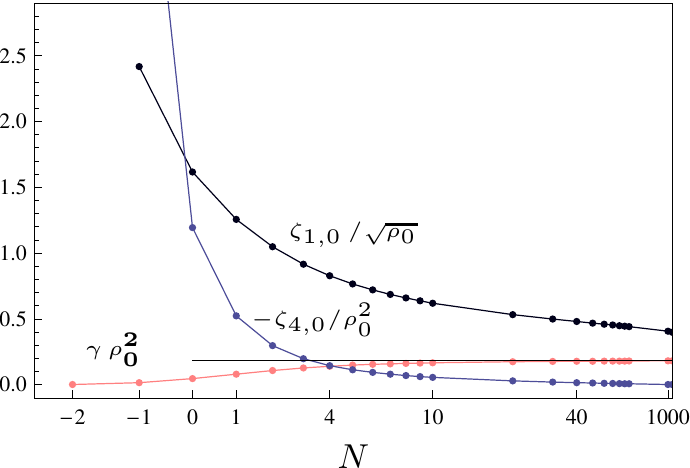}
\end{center}
\vskip-2mm
\caption{The $N$-dependence for the expansion coefficients $\gamma$, $\zeta_{1,0}$ and $\zeta_{4,0}$ at the Wilson-Fisher fixed point in units of $\rho_0(N)$, see Tab.~\ref{tab:asymptcoeffs}. The black horizontal line 
indicates the infinite-$N$ limit for $\gamma$.
	}\label{fig:asympt_N}
\end{figure}

In Tab.~\ref{tab:asymptcoeffs}, we present our results for the expansion coefficients 
\eq{unknowns}, as obtained from the numerical integration of the flow 
equation, using the 
techniques discussed in Sec.~\ref{CS}. For the determination of the large field 
coefficient $\gamma$, we use an asymptotic expansion \eq{eqn:largerho} up 
to order $\rho^{-30}$. For the coefficients $\zeta_{1,0}$ and $\zeta_{4,0}$ we use an 
asymptotic expansion up to order $(\sqrt{-\rho})^{-14}$, corresponding to about a hundred 
terms in the series \eq{eq:negasympt}. Note also that we have rescaled the coefficients 
\eq{unknowns} with $\rho_0(N)$ to factor-out redundant 
$N$-dependences. Again, this is equivalent to a rescaling with \eq{A} of \eq{eq:flowprime} for the choice $A=\rho_0(N)$.
The plot in Fig.~\ref{fig:asympt_N} illustrates that in this normalisation
the coefficient $\gamma(N)$  smoothly approaches its infinte-$N$ asymptotic 
value given by 
\begin{equation}
{\gamma(N)}\times{\rho_0(N)^2}=\frac{16}{9\pi^2}\approx 0.180\,126\cdots\,,
\end{equation}
in accord with the findings in \cite{Litim:2016hlb}. 

\begin{table*}
\begin{center}
\begin{tabular}{cllll ||  cllll}
\toprule
\rowcolor{LightGreen}
 &&&&&&&&&\\[-3mm]
\rowcolor{LightGreen}
$\ \bm N\ $&
$\ \ \ \bm{\gamma\cdot\rho_0^2}$&
$\quad\ \bm{\zeta_{1,0}^2}$&
$\ \bm{\zeta_{1,0}/\sqrt{\rho_0}}$&
$\ \bm{\zeta_{4,0}/\rho_0}$
&$\ \bm  N\ $&
$\ \ \bm{\gamma\cdot\rho_0^2}$&
$\quad\bm{\zeta_{1,0}^2}$&
$\ \bm{\zeta_{1,0}/\sqrt{\rho_0}}$&
$\ \bm{\zeta_{4,0}/\rho_0}$ \\[.8mm]
\midrule
&&&&\\[-3mm]
$-2$&\ \ \ \ n.a.&3.28126470& \ \ \ \ \ n.a. &\ \ \ \ n.a. &10&0.1653897&3.8694520&0.6186249&0.552110	\\
\rowcolor{LightGray}
$-1$&0.0145616&3.09527890&2.41639530&2.622610	&20&0.1732571&5.6586330&0.5312423&0.5555058	\\
 0&0.0457930&2.95035660&1.61520140&1.34971010	&30&0.175657&7.4389030&0.4976890&0.5648128	\\
\rowcolor{LightGray}
 1&0.0790721&2.86082720&1.25550850&0.94716139	&40&0.176816&9.1926050&0.479246&0.57232897	\\
 2&0.1064012&2.83512060&1.04747710&0.7655842	&50&0.1774976&10.9247910&0.467346&0.578200229	\\
\rowcolor{LightGray}
 3&0.125896&2.87007160&0.9152994&0.6724176	&60&0.177946&12.640250&0.458928&0.582876187	\\
 4&0.138984&2.95328100&0.8267318&0.6214516	&70&0.178264&14.342490&0.452607&0.586688428	\\
\rowcolor{LightGray}
 5&0.147731&3.07041660&0.7646765&0.59231311	&80&0.178501566&16.034030&0.447656&0.589862023	\\
 6&0.1537283&3.2099170&0.7193494&0.5750751	&90&0.1786850&17.716710&0.443654&0.592551002	\\
\rowcolor{LightGray}
 7&0.1579893&3.3638350&0.6849811&0.564635	&$10^2$&0.1788313&19.391930&0.440342&0.59486339	\\
 8&0.161128&3.5270400&0.6580723&0.558249	&$10^3$&0.1799987&164.16480&0.405172&0.62464	\\
\rowcolor{LightGray}
 9&0.1635185&3.6962730&0.6364287&0.554376	&$10^4$&0.180113&1582.0140&0.39774&0.63225	\\
\bottomrule
\end{tabular}
\end{center}\caption{Numerical results for the asymptotic expansion coefficients \eq{unknowns} for all $N$ ($\rho_0$ is
given in Tab.~\ref{tab:results_couplings}). 
We quote only those digits which remain stable under variations of the matching points and under a variation of the  corresponding approximation order by $\Delta M=2$.}\label{tab:asymptcoeffs}
\end{table*}

Similarly, for large negative $\rho$, corresponding to purely imaginary fields and assuming that $\rho_0$ is the only global ``scale'' of the fixed point solution, naive dimensional analysis suggests that the coefficients $\zeta_{1,0}(N)$ scale proportionally to $\sqrt{\rho_0}(N)$, and that $\zeta_{4,0}(N)$ should scale proportionally to $\rho_0^2(N)$, modulo logarithmic corrections.
Our results for these are shown in Tab.~\ref{tab:asymptcoeffs} and in Fig.~\ref{fig:asympt_N}. 
The expected behaviour is confirmed for $\zeta_{1,0}(N)$. Unlike the coefficient $\gamma(N)$, however, it has not yet fully settled to its asymptotic value. The  reason for this is that the limit $N\to \infty$ is continuous only for real fields with $\rho>\rho_c$, but not so for purely imaginary fields with $\rho <\rho_c$, see Sec.~\ref{mass}.  In fact, the asymptotic solution for large negative $\rho$ at $N=\infty$ is different from the expansion at any finite $N$ due to the absence of the radial mode fluctuations.  We also find that $\zeta_{4,0}(N)$ roughly scales as $\propto \rho_0(N)$, rather than $\propto \rho_0^2(N)$. For large-$N$, it is therefore smaller by a power of $N$ than naively expected. This behaviour indicates that subleading terms are  more strongly suppressed than the leading one, for large $N$, in accord with  \eq{convexhull}.

 \subsection{Radii of convergence}
\begin{figure}[t]
 \includegraphics[width=.8\hsize]{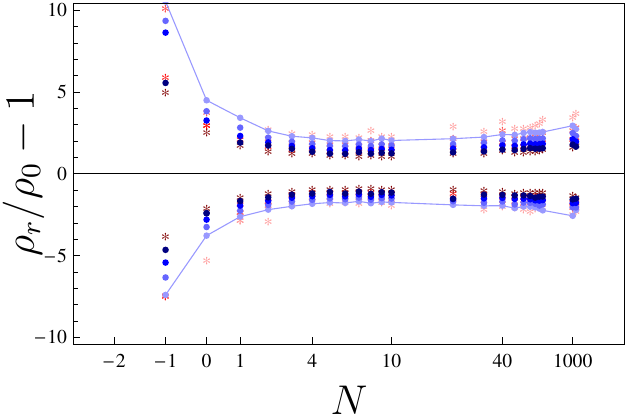}
\vskip1mm 
\caption{The $N$-dependence of the radius of convergence for the polynomial 
 	expansion around the  potential minimum. Stars (circles) represent results 
	at approximation order $M=20$ ($M=30$); see main text.}\label{fig:radius}
\end{figure}

In order to further quanitfy the above we 
define an empirical radius of convergence for the small-field 
expansion around $\rho=\rho_0$ based on the data, see Fig.~\ref{fig:radius}. To that end, we determine a maximal (minimal) field value
$\rho_r$ ($\rho_m$) such that the accuracy of the fixed point solution $N_{\rm acc}(\rho)$, defined in \eq{Nacc}, remains larger than a pre-set desired accuracy $N_{\rm da}$ for all fields $\rho_0<\rho<\rho_r$ ($\rho_m<\rho<\rho_0$).

Our results are illustrated in Fig.~\ref{fig:radius}. 
The blue circles (red stars)
represent for each value of $N$ the results for $\rho_r$ (upper lines) and $\rho_m$ (lower lines),
for a polynomial approximation at order  $M=30$ ($M=20$). The colour 
intensity increases with increasing ``desired accuracy'' $N_{\rm da}=0,2,5,6$. 
The data points above and below the 0-axis correspond to the radius of 
convergence above and below the potential minimum, respectively, indicating 
approximately symmetric convergence properties in these two domains,  $\rho_r-\rho_0\approx\rho_0-\rho_m$. The common radius of convergence is read off as 
\begin{equation}
R_A=\rho_r-\rho_0\,.
\end{equation}
In order 
to guide the eye we have connected  those data points for $M=30$ where the 
difference between the full result and  the polynomial approximation in terms of $N_{\rm acc}$ is of order one; this line is our estimate for the radius of convergence. 
Broadly speaking, we find that
\begin{equation}
R_A\approx (2- 3)\,\rho_0
\end{equation}
for  $N>1$. It is intuitively clear that the 
dark points lie closer to the origin. In fact, the requirement that $N_{\rm acc}=6$ is a much more 
stringent choice for the definition of the radius of convergence. 
This empirical definition for the radius of convergence 
for intermediate values of $N$ is compatible with \eq{eq:R}.
The results for $M=20$, marked by red stars, behave in a very similary way.
In fact, the results agree very well over a wide range of $N$ indicating that 
an excellent accuracy is already achieved at $M=20$ irrespective of the universality class.

\begin{figure*}
\begin{center}
\includegraphics[width=.99\hsize]{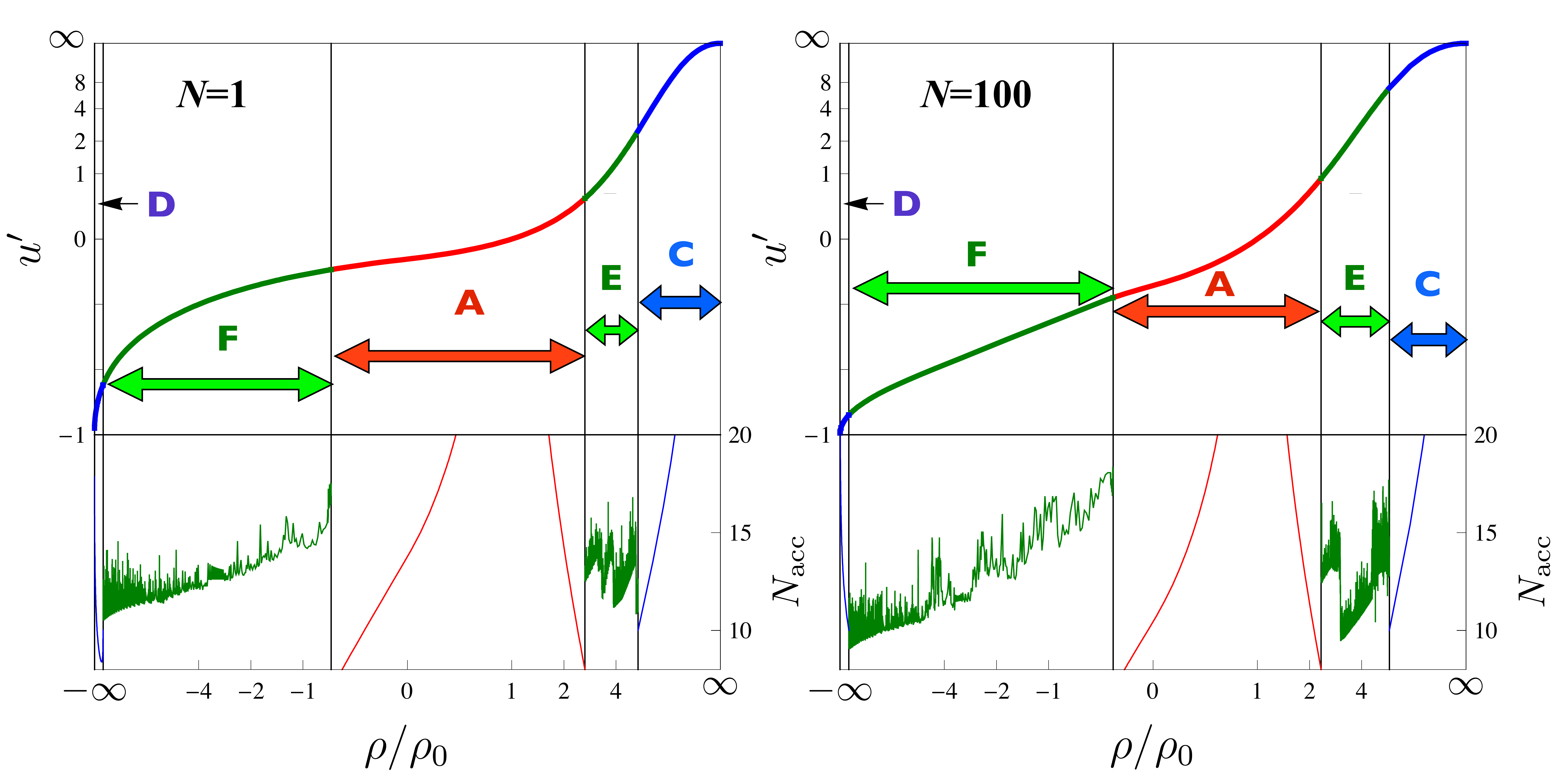}
\vskip-3cm
\end{center}
\caption{Accuracy of the global Wilson-Fisher fixed point $u'_*$  in field space, comparing analytical expansion schemes and numerical integration for the universality classes $N=1$ (left panel) and $N=100$ (right panel). In each diagram, the upper panels show the first derivative of the full global solution $u'(\rho)$ and indicates the regimes where 
 a small field expansion $(A)$, large real field expansion $(C)$,
 large imaginary field expansion $(D)$, or intermediate
 numerical integrations for real $(E)$ and imaginary fields $(F)$ have been used.  The lower panels indicate the numerical accuracy of the fixed point solution in terms of the number of accurate decimal places $N_{\rm acc}(\rho)$  \eq{Nacc} for the various regimes. The axes are rescaled as in Fig.~\ref{fig:potential1} for better display.
 }\label{fig:prec}
\end{figure*}

Exemplarily, we evaluate the radius of convergence with higher accuracy for the Ising universality class $N=1$, and for the infinite-$N$ limit. In the Ising case, we use the Mercer-Roberts test \cite{Litim:2016hlb}  for the numerical expansion coefficients 
to find the estimate
\begin{equation}\label{Radius1}
\frac{R_A}{\rho_0}=2.7(74)
\end{equation}
for $N=1$, which is in good agreement with our findings in Fig.~\ref{fig:radius} and the estimate of the radius based on the present study, displayed in Fig.~\ref{fig:prec}. 

At infinite $N$, the radius of convergence is related to the unique pair of complex conjugate solutions $z_p$ and $z_p^*$ of the transcendental equation \cite{Litim:2016hlb}
\begin{equation}\label{root}
H'\left(H^{-1}(z_p)\right)=0\,,
\end{equation}
where the function $H$ is given in \eq{H} and $H^{-1}$ denotes its  inverse, $H^{-1}\left(H(z)\right)=z$. From \eq{root}, one then determines the radius of convergence in units of the vacuum expectation value as ${R_A}=|z_p|\times {\rho_0}$. Quantitatively,
\begin{equation}
\frac{R_A}{\rho_0}=3.183\,547\,200\,
\cdots\,,
\end{equation}
which is in good agreement with the numerical estimates achieved for large $N$, see Fig.~\ref{fig:radius}.

\subsection{Accuracy of global fixed point}

\begin{figure*}
\begin{center}
\includegraphics[width=.5\hsize]{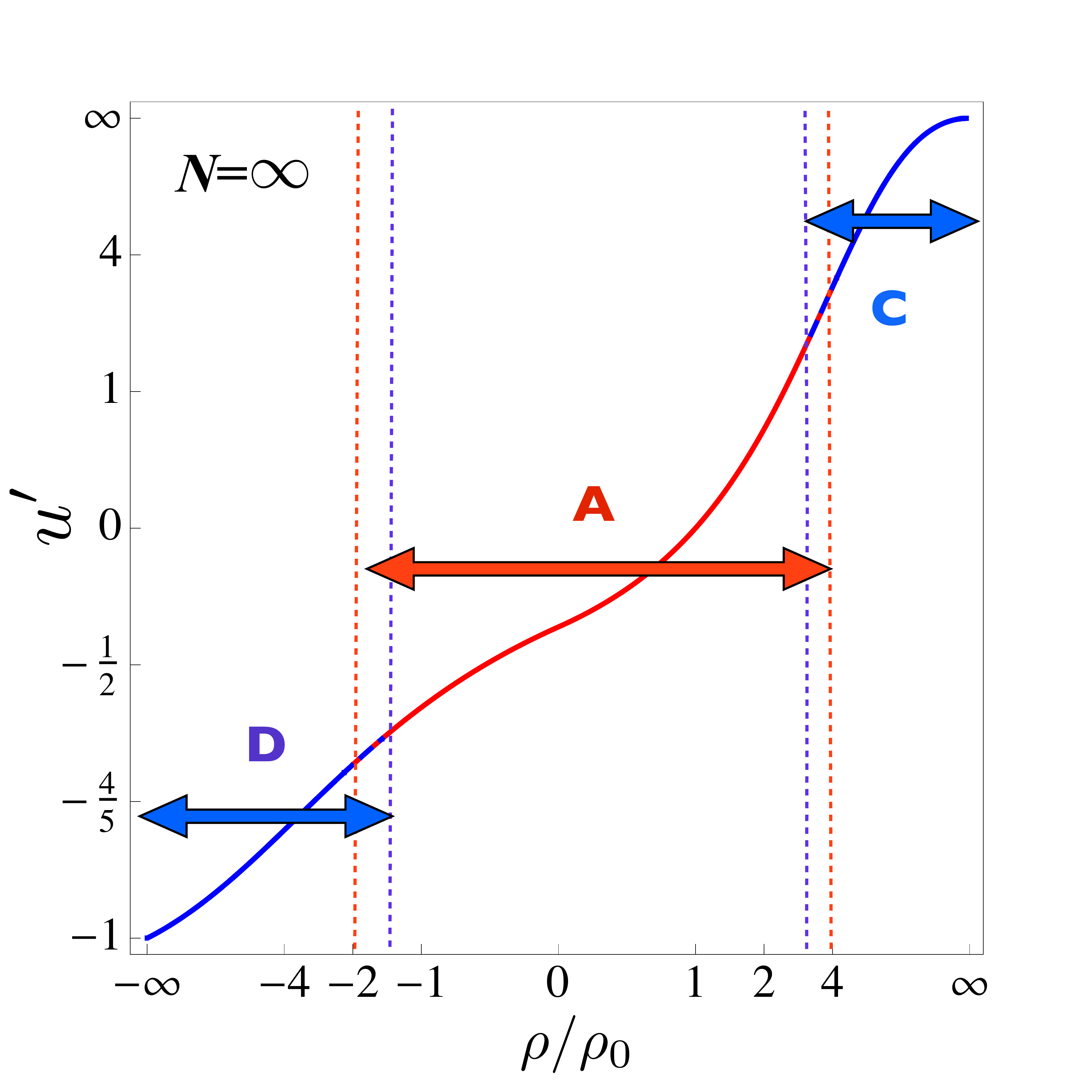}
\end{center}
\vskip-.5cm
\caption{The global Wilson-Fisher fixed point $u'_*(\rho)$ in the infinite $N$ limit, showing that the  small field expansion $(A)$, the large real field expansion $(C)$ and the 
 large imaginary field expansion $(D)$ have overlapping radii of convergence.
Axes are rescaled as in Fig.~\ref{fig:potential1} for better display.
 }\label{fig:inf}
 \end{figure*}

In Fig.~\ref{fig:prec} we illustrate exemplarily for $N=1$ and $N=100$
the accuracy of the fixed point solution to leading order in the derivative expansion, covering the entire field space $\rho\in [-\infty,\infty]$ of real and purely imaginary fields. We also highlight the regime where we used a small field expansion around the potential minimum (region $A$), asymptotic expansions  for large positive  $\rho$ (region $C$), large negative $\rho$ (region $D$), and the intermediate regions  connecting these by a full numerical integration (regions $E$ and $F$).
The upper panel shows  $u^\prime(\rho)$ (left vertical axis), while the lower panel provides the accuracy $N_{\rm acc}^{}(\rho)$ \eq{Nacc} (right vertical axis) of the fixed point. These findings should be compared with  Fig.~\ref{fig:inf}, where we show the exact global Wilson-Fisher fixed point at infinite $N$, including the radii of convergence of local expansions. 
We make the following observations:

$(i)$ At infinite $N$, only Goldstone mode fluctuations contribute and the radial mode has decoupled. We observe that the local expansion $A$ has an overlapping radius of convergence with those of expansions $C$ and $D$, see Fig.~\ref{fig:inf}. Consequently, the parameters $m^2, \gamma$ and $\xi$ are  uniquely determined through the couplings of expansion $A$, see Tab.~\ref{tParameters}. This provides us with a global and  analytical solution $u'_*(\rho)$  for all fields \cite{Litim:2016hlb}.

$(ii)$ At finite $N$, 
the fluctuations of the radial mode contribute. 
We observe that large and small field expansions no longer have  overlapping radii of convergence, illustrated in Fig.~\ref{fig:prec}  through the intermediate (green) regions $E$ and $F$. 
The gap in field space is narrower for real  $(E)$ than for imaginary fields $(F)$, and mildly depends on $N$ and the competition between Goldstone and radial mode fluctuations. The fixed point equation, albeit stiff, admits a stable numerical  integration in the intermediate regimes $E$ and $F$. The accuracy in region $E$ is rather homogeneous. In region $F$ (where the integration starts at large negative $\rho$) the accuracy slowly decreases inwards, with decreasing $-\rho$. In either case the overall accuracy is solely limited by the accuracy of the adopted numerical algorithm.\footnote{Semi-analytical ideas to bridge the gap using conformal mapping are discussed in \cite{Abbasbandy:2011ij}.}   

$(iii)$ The accuracy of all local expansions decreases 
with increasing distance from the expansion point, as is evident from
the decrease of  $N_{\rm acc}$ in the bottom-half of the plots, see Fig.~\ref{fig:prec}. 
This is a necessary fingerprint of local expansions.
 We have checked that local expansions agree with the numerical integration in the regions where they overlap. This way, an overall accuracy of $10-20$ relevant digits is achieved.

\subsection{Accuracy of universal exponents}

\begin{figure*}
  \begin{center}
\includegraphics[width=.95\hsize]{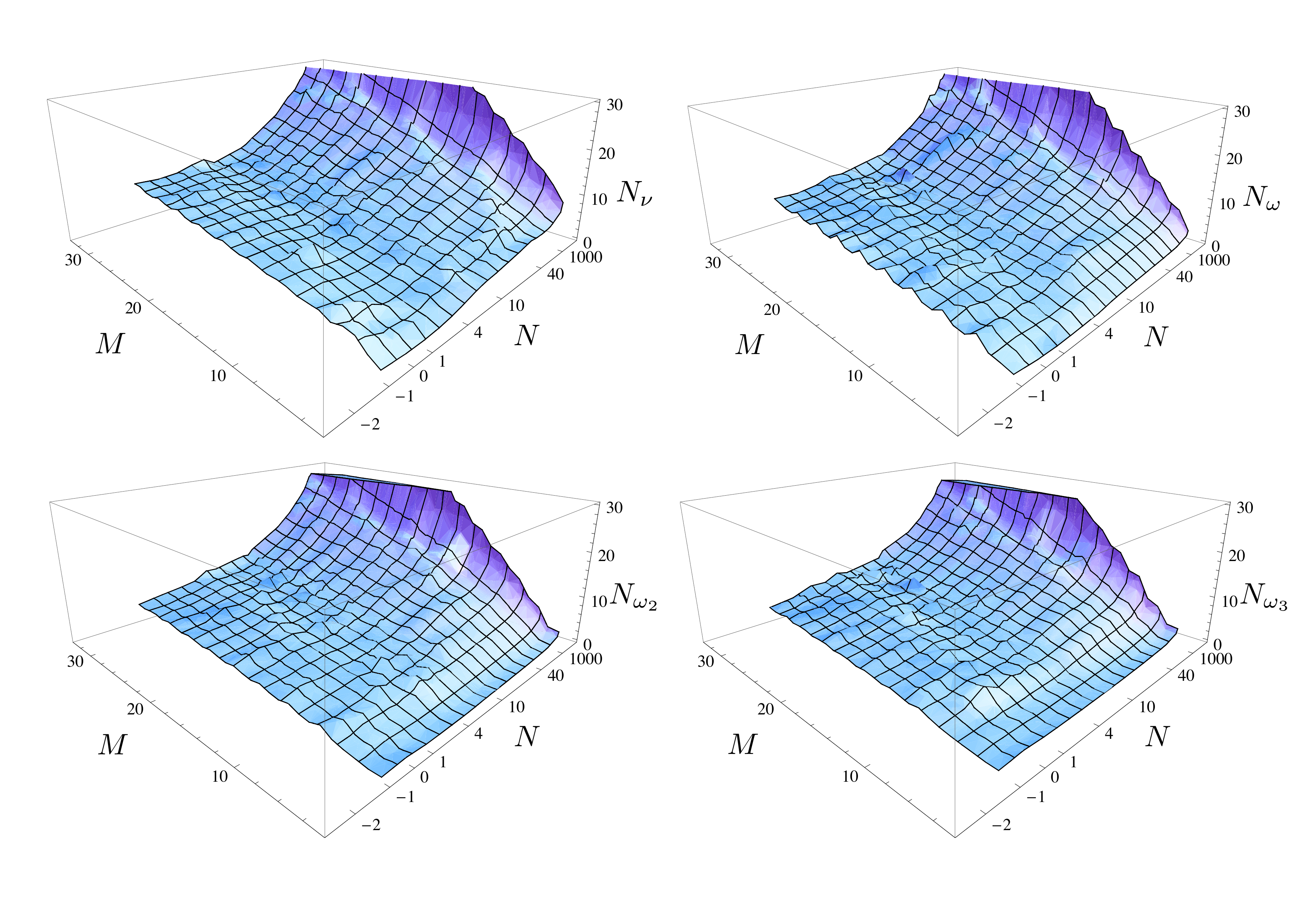}
\vskip-.75cm  \caption{Accuracy of scaling exponents, characterised by the number of stable relevant digits $N_X$ \eq{NX} for the exponents $X=\nu, \omega, \omega_1, \omega_2$ (from top left  to bottom right, respectively) for all universality classes studied here, and as a function of the order of the polynomial approximation $M$. We note that the accuracy at fixed order in the approximation increases rapidly with increasing $N$ (see text).}
  \label{fig:error_Nnu}
  \end{center}
\end{figure*}

In Fig.~\ref{fig:error_Nnu}, the accuracy and convergence  to leading order in the derivative expansion
of the four leading scaling exponents are shown as a surface plot for all 
$N$ and all approximation orders $M$. 
All plots show a monotonic increase in the number of significant digits ${N_{X_n}}$
as the order of the polynomial approximation is increased, $M\to M+1$. This means that 
the scaling potential can indeed be improved systematically, and for all $N$. 
Interestingly, the accuracy in the result is systematically higher for larger 
values of $N$. This result establishes that the numerical strategy used here is applicable, and that the boundary condition \eq{boundary} is self-consistent. We have also found indications that an improved boundary condition, informed by semi-analytical input for the higher order couplings, improves the convergence even further. Given the quality in the results, we have not attempted to implement more refined strategies here.

\subsection{Wilson-Polchinski flow}
We finally discuss a link between  optimised and Wilson-Polchinski flows related to singularities of their fixed point solutions in the regime of purely imaginary fields. In~\cite{Litim:2005us}, it has been conjectured that the optimised RG flows~\cite{Litim:2001up,Litim:2000ci} studied here are equivalent to the Wilson-Polchinski RG flow~\cite{Wilson:1973jj,Polchinski:1983gv} in the same approximation. In fact, both of these RG flows are related by a Legendre transform. By now, their equivalence has been established both numerically \cite{Litim:2005us,Bervillier:2007rc}, by evaluating universal scaling exponents in either version,  and analytically~\cite{Morris:2005ck,Bervillier:2007rc}, with the help of the Legendre transform. 

Here, we briefly discuss how the singularity for large and purely imaginary fields translates to the corresponding fixed point solution of the Wilson-Polchinski flow. 
For asympotically large imaginary fields  the first 
derivative of the scaling potential  takes the form~(\ref{eq:negasympt}),
\begin{equation}\label{optPole}
	u_*'(\rho) = -1 + \frac{\zeta}{\sqrt{-\rho}} +{\rm subleading}\,,
\end{equation}
up to next-to-next-to-leading order  corrections, see \eq{eq:negasympt}, \eq{zetasDef}.  We note that since the variable $\rho\propto \phi_a\phi^a$ takes all real values on the fixed point solution, the fields $\phi^a$ exhaust all values on the imaginary field axis. In particular,  $u'_*$ remains finite throughout.

We now turn to the Wilson-Polchinski flow for the potential $v(\rho)$. In $d=3$ dimensions and for $N\neq 0$ it can be written as  \cite{Litim:2005us}
\begin{equation}\label{Polchinski}
\partial_t v-3\,v + \rho\, v' = v' + \frac{2}{N} \,\rho\,v''-2\,\rho\,(v')^2,
\end{equation}
where $\rho=\frac 12 \phi^a\phi_a/k$ is the  square of the field $\phi_a$ in units of the RG scale $k$. At the Wilson-Fisher fixed point $v_*(\rho)$, and in the regime of purely imaginary fields and negative $\rho$, it has been observed that~\cite{Osborn:2009vs} 
\begin{equation}\label{poleWP}
	v_*'(\rho) \sim - {1 \over  \rho + \rho_p}+{\rm subleading}\,,
\end{equation} 
where the parameter $\rho_p(N)>0$ depends on the universality class $N$. In contrast to \eq{optPole}, the result \eq{poleWP} shows that the integrated Wilson-Polchinski flow exhibits a simple pole at finite fields  
\beq \varphi=\pm i \sqrt{2\rho_p}
\eeq 
on  the imaginary field axis, with $\varphi=|\phi^a|$. Performing a Legendre transform following \cite{Bervillier:2007rc} provides us with relations between the effective potential $v(\rho)$ of the Wilson-Polchinski flow and the potential $u(\rho)$ from the optimised RG. This leaves us with an explicit link between the parameters $\rho_p$ in \eq{poleWP} and the coefficient $\zeta$ in \eq{optPole},
\begin{equation}\label{zeta}
\rho_p=\zeta^2\,.
\end{equation}
Hence, although the singularities in the complexified field plane in either setting
are qualitatively different, as evidenced in \eq{optPole}, \eq{poleWP}, 
the behaviour of the Wilson-Fisher fixed point for asymptotically large and purely imaginary fields in one setting dictates the location of a pole in the other,
both given by the same parameter $\zeta$. 
In Tab.~\ref{tab:Osborn} we confirm the result \eq{zeta} numerically, comparing our results for $\zeta$ from \eq{optPole} with those derived numerically from \eq{poleWP} in~\cite{Osborn:2009vs}. Besides offering a  consistency check for the analytical  results 
of \cite{Litim:2016hlb},
the quantitative agreement  also confirms equivalence of renormalisation group flows even for purely imaginary fields.

\begin{table*}[t]
\begin{tabular}{cLlLlLc}
\rowcolor{LightBlue}
\toprule
\cellcolor{LightBlue}
&&&&&&
\cellcolor{LightBlue}
\\[-3mm]
\rowcolor{LightBlue}
\cellcolor{LightBlue}
$\bm\zeta^{\bm 2}$&
\multicolumn{1}{c}{$N=1$}&
\multicolumn{1}{c}{$N=2$}&
\multicolumn{1}{c}{$N=3$}&
\multicolumn{1}{c}{$N=4$}&
\ \ {$N=10$}&
\cellcolor{LightBlue}
\bf info\\
\midrule
\cellcolor{LightGreen}&&&&&&\cellcolor{LightGreen}\\[-3mm]
\cellcolor{LightGreen}
${}\quad$Wilson-Polchinski RG$\quad$  &\ 2.862& \ 2.836& \ 2.871&\  2.954&\ 3.871&\cellcolor{LightGreen}Ref.~\cite{Osborn:2009vs}\\
\cellcolor{LightGreen}
optimised RG   &\ 2.86082720\ &\ 2.83512060\ &\ 2.87007160\ &\ 2.95328100\ &\ 3.8694520\ &\cellcolor{LightGreen}\ this work\         \\[1mm]
\bottomrule
\end{tabular}
\caption{Comparison of the parameter $\zeta^2$ for various $O(N)$ universality classes as obtained from the Wilson-Polchinski RG \cite{Osborn:2009vs}  and the optimised RG (this work).}\label{tab:Osborn}
\end{table*}

\section{\bf Discussion}\label{sec:Conclusion}
The combined use of analytical and  numerical methods is often vital for the understanding of non-perturbative phenomena in quantum and statistical field theory.  Here, in combination with the analytical studies put forward in \cite{Litim:2016hlb}, we have performed a detailed  numerical analysis of $O(N)$ symmetric scalar field theories at their Wilson-Fisher fixed point in three dimensions. We have exploited polynomial expansions up to very high order, showing that these deliver reliable and rapidly converging results for scaling exponents when combined with  expansions about asymptotically large, or purely imaginary fields, and numerical integration. In this manner we have achieved  global fixed point solutions for the quantum effective action with moderate numerical effort, also obtaining high numerical accuracy for scaling exponents to leading order in the derivative expansion, critical couplings, field ratios, and a number of other universal, super-universal, and non-universal features of scaling solutions. Our findings  strengthen the view that scaling exponents can efficiently be deduced from  data encoded in the small field region  of the fixed point effective action \cite{Litim:2002cf}.

Furthermore, our study showed that large- and small-field expansions often do not yield overlapping radii of convergence owing  largely to the fluctuations of the radial mode.
This affects the physically most interesting universality classes where the number of fields, parametrised by $N$, is  finite and small. Interestingly, this  phenomenon is absent at infinite $N$ where the radial mode decouples and  the Goldstone modes dominate instead.  The competition between longitudinal and transversal fluctuations appears to be most pronounced for moderate values of $N$ where the Wilson-Fisher fixed point is more strongly coupled. Quantitatively, however, we found that the non-perturbative gap in field space  is narrower for real than for purely imaginary fields, and that the intermediate field region is very efficiently covered by straightforward and stable numerical integration. The procedure appears to be  sufficient for all technical purposes. 

We have also provided a detailed analysis covering the entire field space including the region of purely imaginary fields. In the latter region, qualitative differences between longitudinal and transversal mode fluctuations are particularly pronounced. The strict convexity bound \eq{convexity} is relaxed for the radial modes at infinite $N$ without causing a singularity, but not so at any finite $N$ where the  bound remains firmly in place. Consequently, the $N$-dependence of the effective potential is discontinuous at infinite $N$.
No such discontinuity is visible in the  region where fields are real.
As a by-product, we  also observed that while fixed point solutions of \eq{eq:flow} are free of singularities for any real or purely imaginary field, those of the Polchinski flow \eq{Polchinski} unavoidably develop singularities at finite and purely imaginary field. Interestingly, the behaviour of the Wilson-Fisher fixed point solution for large and purely imaginary fields in the former setting  dictates the location of the singularity in the latter,  for any universality class. Overall, we have thus established a comprehensive global picture covering all the salient features of non-perturbative scaling solutions. 

Although we have limited ourselves to the leading order in the derivative expansion, our line of reasoning and methodology can be used beyond the leading order as well as for other systematic expansions.
At second and fourth order in the derivative expansion, some focus has been given to optimised choices for the Wilsonian momentum cutoff
\cite{Canet:2002gs,Canet:2003qd,Canet:2004xe,Litim:2000ci,Litim:2010tt}. 
It is conceivable that predictions for  universal exponents can be further improved by exploiting the structure of equations in the complexified field plane \cite{Litim:2016hlb}, possibly supported by conformal mappings \cite{Abbasbandy:2011ij} and the close links with the Polchinski flow. Some of the qualitative and quantitative insights of our study will also prove useful for fixed point searches  in other theories, e.g.~$3d$ Wess-Zumino models \cite{Litim:2011bf,Heilmann:2012yf}, or $4d$ quantum gravity \cite{Falls:2014tra}. While the gravitational renormalisation group is  more involved than the equations explored here, the interplay between singularities in the complexified field plane and radii of convergence is operative  in gravity as well \cite{Falls:2016wsa}.

\section*{\bf Acknowledgements}
We thank Hugh Osborn for discussions and CERN for providing computing time.
This work is supported by the 
Science and Technology Facilities Council 
under grant number ST/L000504/1, and by the European Research Council under the European 
Community's Seventh Framework Programme FP7 ERC grant agreement No 279757.

\bibliographystyle{JHEP}
\bibliography{biblio2_modif}

\end{document}